\documentclass[preprint,12pt]{aastex}

\def \eg{{e.g.,}}

\def \etal{{et~al.\null}}
\def \ie{{i.e.,}}

\def \h7{{h_{70}}}

\title{Ly$\alpha$ Emission-Line Galaxies 
at $z = 3.1$ in the Extended Chandra Deep Field South}

\begin{document}

\shorttitle{Ly$\alpha$ Emission-Line Galaxies}

\author{Caryl Gronwall, Robin Ciardullo, Thomas Hickey}
\affil{Department of Astronomy \& Astrophysics, The Pennsylvania State
University \\ 525 Davey Lab, University Park, PA 16802}
\email{caryl@astro.psu.edu, rbc@astro.psu.edu, tomhickey@astro.psu.edu}

\author{Eric Gawiser\altaffilmark{1}}
\affil{Yale Astronomy Department and Yale Center for Astronomy and 
Astrophysics \\ Yale University, P.O. Box 208121, New Haven, CT 06520}
\email{gawiser@astro.yale.edu}

\author{John J. Feldmeier\altaffilmark{1}}
\affil{Department of Physics \& Astronomy, Youngstown State University,
Youngstown, OH 44555-2001}
\email{jjfeldmeier@ysu.edu}

\author{Pieter G. van Dokkum, C. Megan Urry, David Herrera}
\affil{Yale Astronomy Department and Yale Center for Astronomy and 
Astrophysics and Yale Physics Department \\ Yale University, P.O. Box 208121, 
New Haven, CT 06520}
\email{pieter.vandokkum@yale.edu, meg.urry@yale.edu, david.herrera@yale.edu}

\author{Bret D. Lehmer}
\affil{Department of Astronomy \& Astrophysics, The Pennsylvania State
University \\ 525 Davey Lab, University Park, PA 16802}
\email{blehmer@astro.psu.edu}

\author{Leopoldo Infante, Alvaro Orsi}
\affil{Departmento de Astronom\'ia y Astrof\'isica, Pontificia Universidad 
Cat\'olica de Chile, Casilla 306, Santiago 22, Chile}
\email{linfante@astro.puc.cl, aaorsi@astro.puc.cl}

\author{Danilo Marchesini}
\affil{Yale Astronomy Department and Yale Center for Astronomy and 
Astrophysics \\ Yale University, P.O. Box 208121, New Haven, CT 06520}
\email{danilom@astro.yale.edu}

\author{Guillermo A. Blanc}
\affil{Astronomy Department, University of Texas, Austin, TX 78712}
\email{gblancm@astro.as.utexas.edu}

\author{Harold Francke, Paulina Lira}
\affil{Departamento de Astronom\'ia, Universidad de Chile,
Casilla 36-D, Santiago, Chile}
\email{hfrancke@das.uchile.cl, plira@das.uchile.cl}

\and
\author{Ezequiel Treister}
\affil{European Southern Observatory, Casilla 19001, Santiago, Chile}
\email{etreiste@eso.org}

\altaffiltext{1} {NSF Astronomy and Astrophysics Postdoctoral Fellow}
 
\begin{abstract}
We describe the results of an extremely deep, 0.28~deg$^2$ survey for
$z = 3.1$ Ly$\alpha$ emission-line galaxies in the Extended Chandra 
Deep Field South.   By using a narrow-band 5000~\AA\ filter and complementary
broadband photometry from the MUSYC survey,  we identify a statistically
complete sample of 162 galaxies with monochromatic fluxes brighter than
$1.5 \times 10^{-17}$~ergs~cm$^{-2}$~s$^{-1}$ and observers frame equivalent
widths greater than 80~\AA\null.    We show that the equivalent width
distribution of these objects follows an exponential with a rest-frame scale 
length of $w_0 = 76^{+11}_{-8}$~\AA.  In addition, we show that in the emission 
line, the luminosity function of Ly$\alpha$ galaxies has a faint-end power-law 
slope of $\alpha = -1.49^{+0.45}_{-0.34}$, a bright-end cutoff of 
$\log L^* = 42.64^{+0.26}_{-0.15}$, and a space density above our detection 
thresholds of $1.46 \pm 0.12 \times 10^{-3} \, \h7^{3}$~galaxies~Mpc$^{-3}$.
Finally, by comparing the emission-line and continuum properties of the
LAEs, we show that the star-formation rates derived from Ly$\alpha$
are $\sim 3$ times lower than those inferred from the rest-frame UV continuum.
We use this offset to deduce the existence of a small amount of internal 
extinction within the host galaxies.  This extinction, coupled with the lack 
of extremely-high equivalent width emitters, argues that these galaxies are not
primordial Pop~III objects, though they are young and relatively chemically 
unevolved.

%By examining the continuum properties of our Ly$\alpha$ emitters, we show that
%to $R \sim 25.5$, these objects are $\sim 1/3$ as common as Lyman-break 
%galaxies; by translating their emission-line fluxes into star-formation rates, 
%we show that these galaxies typically form stars at rates between 1 and 
%$10 \, \h7^{-2} \, M_{\odot}$~yr$^{-1}$.  We find that the integrated 
%star-formation rate density for Ly$\alpha$ emitters, in the absence of dust, is
%$6.5^{+5.5}_{-1.0}\times 10^{-3} \, \h7 \, M_{\odot}$~yr$^{-1}$~Mpc$^{-3}$.
%However, we also show that the star-formation rates derived from the 
%Ly$\alpha$ emission-line fluxes are $\sim 3$ times less than those implied 
%from the galaxies' rest-frame UV continua.  We use this offset to infer the 
%existence of a very small amount of internal extinction within the host 
%galaxies, which roughly triples the inferred star formation rate density. 
%This extinction, coupled with the lack of extremely-high equivalent width 
%Ly$\alpha$ emitters, argues that these galaxies are not primordial Pop~III 
%objects, though they are young and relatively chemically unevolved.

\end{abstract}


\section{Introduction}
The past decade has seen an explosion in our ability to detect and study
$z > 3$ galaxies and probe the history of star formation in the universe
\citep[\eg][]{madau}.   This has been mostly due to the development of the 
Lyman-break technique, whereby high redshift galaxies are identified via a
flux discontinuity caused by Lyman-limit absorption \citep[see][]{steidel96a, 
steidel96b}.  By taking deep broadband images, and searching for $U$,
$B$, and $V$-band dropouts, astronomers have been able to explore large-scale 
structure and determine the properties of bright ($L > 0.3 L^*$) galaxies 
between $z \sim 3$ and $z \sim 5$ \citep{giavalisco}.

The stunning success of the Lyman-break technique stands in contrast
to the initial results of Ly$\alpha$ emission-line observations.  The failure
of the first generation of these surveys \citep[\eg][]{depropis, thompson} 
was attributed to internal extinction in the target galaxies \citep{mt81}.
Since Ly$\alpha$ photons are resonantly scattered by interstellar hydrogen,
even a small amount of dust can reduce the emergent emission-line flux 
by several orders of magnitude.  

Fortunately, Ly$\alpha$ surveys
have recently undergone a resurgence.  Starting with the Keck observations of 
\citet{ch98} and \citet{hcm98}, narrow-band searches for Ly$\alpha$ emission 
have been successfully conducted at a number of redshifts, including
$z \sim 2.4$ \citep{stiavelli}, $z \sim 3.1$ \citep{c02, hayashino, venemans,
gawiser}, $z \sim 3.7$ \citep{fujita}, $z \sim 4.5$ \citep{rhoads00}, 
$z \sim 4.9$ \citep{ouchi}, $z \sim 5.7$ \citep{rhoads03, ajiki, tapken}, 
and $z \sim 6.5$ \citep{kodaira, taniguchi}.  The discovery of these 
high-redshift Ly$\alpha$ emitters (LAEs) has opened up a new frontier 
in astronomy.  At $z > 4$, LAEs are as easy to detect than Lyman-break 
galaxies (LBG), and, by  $z > 6$, they are the only galaxies observable from 
the ground.  By selecting galaxies via their Ly$\alpha$ emission, it is 
therefore possible to probe much further down the galaxy continuum
luminosity function than with the Lyman-break technique, and perhaps 
identify the most dust-free objects in the universe.  In addition, by 
using Ly$\alpha$ emitters as tracers of large-scale structure
\citep{steidel00, shimasaku04}, it is possible to efficiently probe
the expansion history of the universe with a minimum of cosmological 
assumptions \citep[\eg][]{blake, seo, koehler}.

Here, we describe the results of a deep survey for Ly$\alpha$ emission-line
galaxies in a 0.28~deg$^2$ region centered on the Extended Chandra Deep 
Field South (ECDF-S).   This region has an extraordinary amount of
complementary data, including high-resolution optical images from the
{\sl Hubble Space Telescope\/} via the Great Observatories Origins Deep 
Survey \citep[GOODS;][]{GOODS} and the Galaxy Evolution from Morphology 
and SEDs program \citep[GEMS;][]{GEMS}, deep groundbased $UBVRIzJHK$ 
photometry from the Multiwavelength Survey by Yale-Chile 
\citep[MUSYC;][]{musyc}, mid- and far-IR observations from {\sl Spitzer,} 
GOODS and MUSYC, and deep X-ray data from {\sl Chandra} \citep{giacconi, 
alexander, lehmer05}.  In Section~2, we describe our observations, which 
include over 28~hours worth of exposures through a narrow-band filter on 
the CTIO 4-m telescope.  We also review the techniques used to detect the 
emission-line galaxies, and discuss the difficulties associated with 
analyzing samples of LAEs discovered via fast-beam instruments.  In 
Section~3, we describe the continuum properties of our Ly$\alpha$ emitters, 
including their rest-frame $m_{1050} - m_{1570}$ colors, and compare their 
space density to that of Lyman-break galaxies.  In Section~4, we examine the 
LAE's equivalent width distribution and show that our sample contains very 
few of the extremely-high equivalent width objects found by
\citet{dawson} at $z = 4.5$.  In Section~5, we present the Ly$\alpha$
emission-line luminosity function, and give values for its best-fit
\citet{schechter} parameters and normalization.  In Section~6, we translate 
these Ly$\alpha$ fluxes into star-formation rates, and consider the properties 
of LAEs in the context of the star-formation rate (SFR) history of
the universe.  We conclude by discussing the implications our 
observations have for surveys aimed at determining cosmic evolution.

For our analysis, we adopt a $\Lambda$CDM cosmology with 
$H_0 = 70$~km~s$^{-1}$~Mpc$^{-1}$ ($\h7 = 1$), $\Omega_M = 0.3$, and 
$\Omega_{\Lambda} = 0.7$.  At $z = 3.1$, this implies a physical scale
of 7.6~kpc per arcsecond.

\section{Observations and Reductions}
\begin{figure}
\plotone{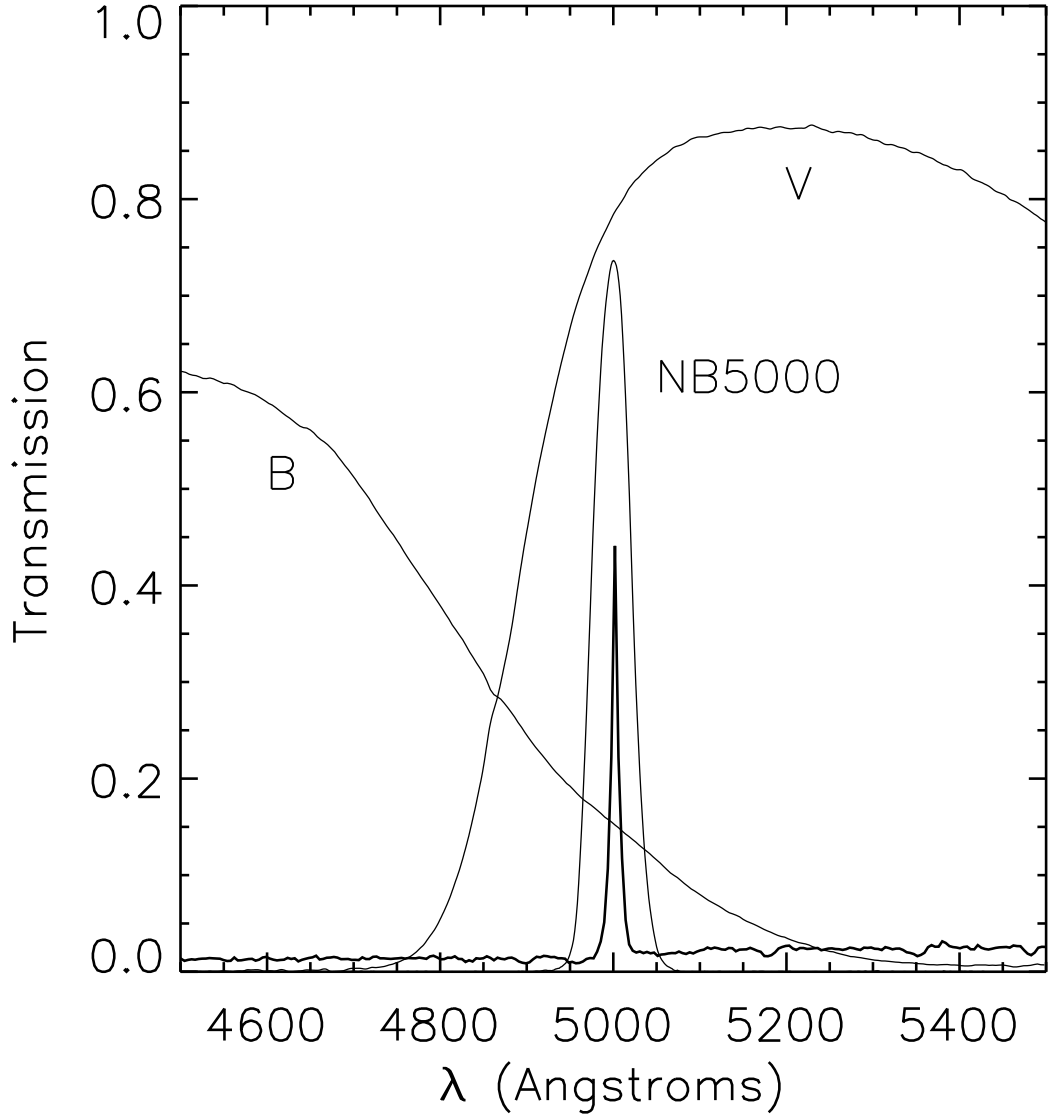}
\figcaption[bandpasses]
{The bandpass of our narrow-band $\lambda 5000$ filter, along with 
those of the $B$ and $V$ filters, which are used to define the 
continuum.  A spectrum of a typical $z = 3.1$ Ly$\alpha$ galaxy is
overlaid for comparison.  Our narrow-band filter isolates the emission
line of Ly$\alpha$ sources with $3.09 \lesssim z \lesssim 3.13$.
\label{bandpasses}
}
\end{figure}

Narrow-band observations of the ECDF-S were performed with the MOSAIC~II
CCD camera on the CTIO Blanco 4-m telescope.  These data consisted of a 
series of 111 exposures taken over 16 nights through a 
50~\AA\ wide full-width-half-maximum (FWHM) $\lambda 5000$ filter (see
Figure~\ref{bandpasses}).  The total exposure time for these images was 
28.17~hr; when the effects of dithering to cover for a dead CCD during
some of the observations are included, the net exposure time becomes 
$\sim 24$~hr.   The total area covered in our survey is 998~arcmin$^{2}$; 
after the regions around bright stars are excluded, this area shrinks
993~arcmin$^{2}$.   The overall seeing on the images is $1\farcs 0$.  
A log of our narrow-band exposures appears in Table~\ref{log}.

\begin{figure}
\figurenum{2}
\plotone{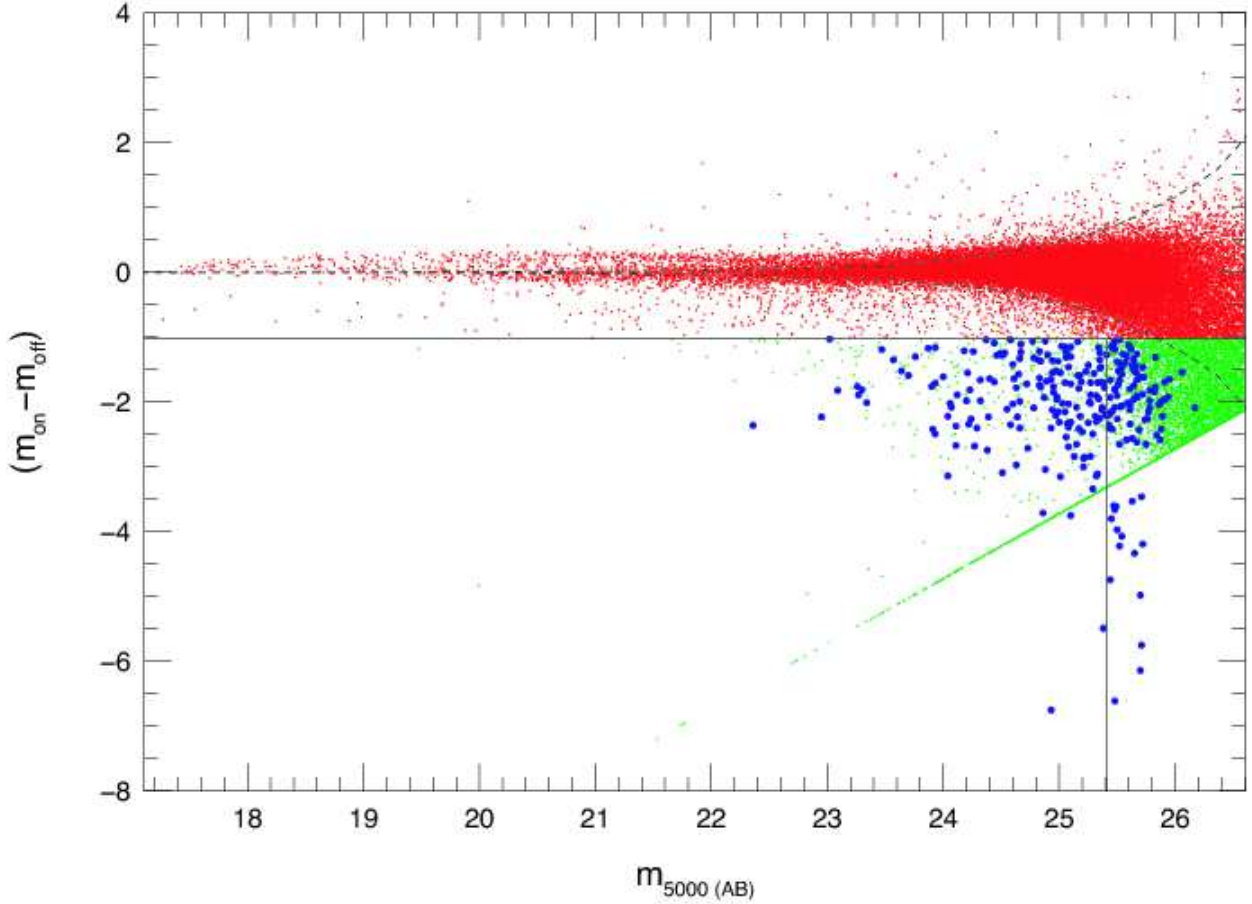}
\figcaption[cmd]
{Excess emission in the narrow-band $\lambda 5000$ filter over the continuum
for objects in our survey field.  The abscissa gives the instrumental $\lambda 
5000$ magnitude, while the ordinate shows the difference between the sources' 
narrow-band and $B+V$ continuum AB magnitudes.  Our narrow-band completeness 
limit of $1.5 \times 10^{-17}$~ergs~cm$^{-2}$~s$^{-1}$ is represented by a
vertical line; our equivalent width limit of 90~\AA\ is shown via the 
horizontal line.  The curve shows the expected $1 \, \sigma$ errors in the 
photometry.  Candidate emission line galaxies are denoted as blue circles;
the green dots indicate LAE candidates found by our detection algorithms, but
rejected upon visual inspection. 
\label{cmd}
} 
\end{figure}

\begin{figure}
\figurenum{3}
\plotone{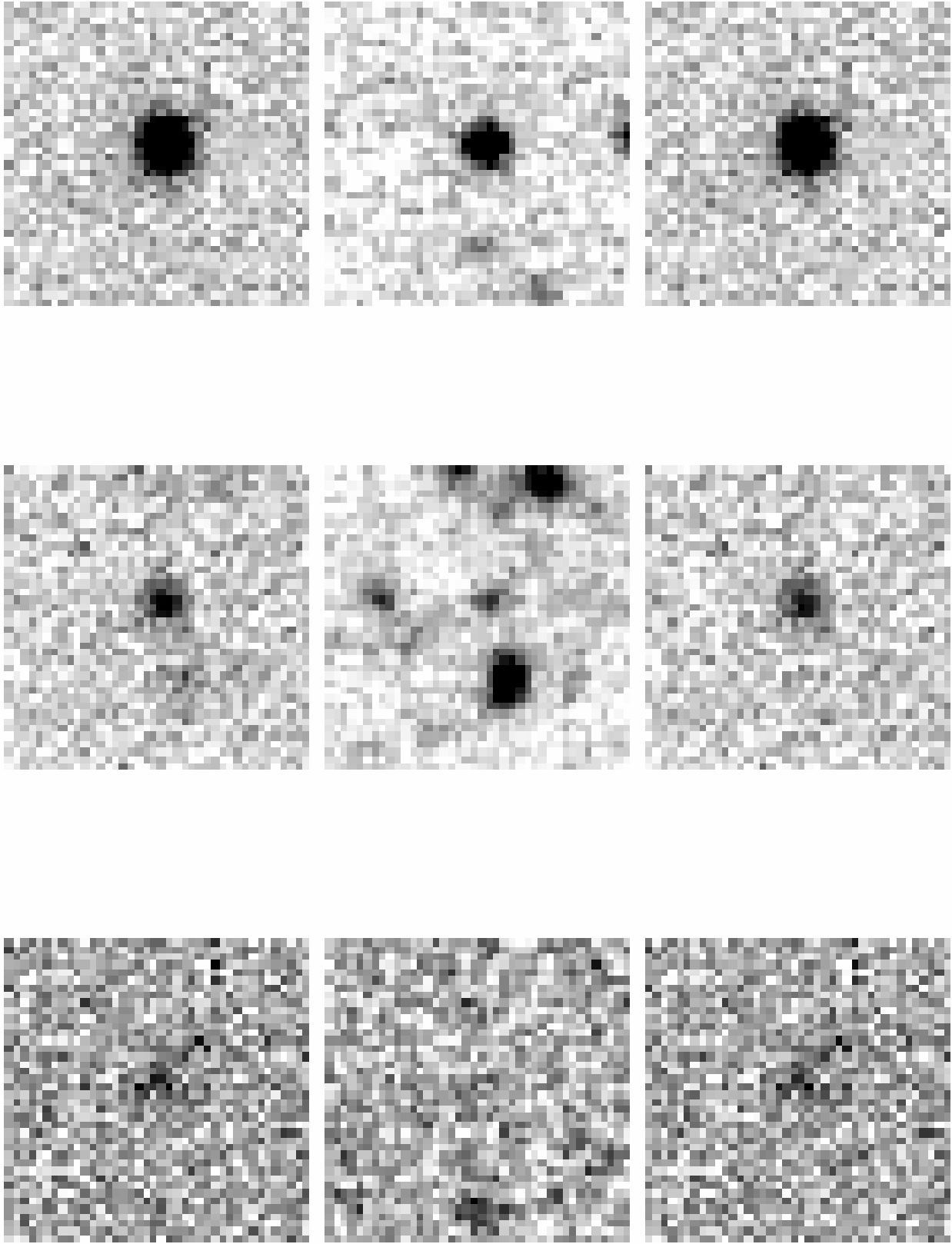}
\figcaption[dimages]
{Narrow-band $\lambda 5000$, $B+V$, and difference images for three
candidate emission-line galaxies.  Each frame is $10\arcsec$ on a side,
with north up and east to the left.  The objects span a range of 
brightness from $\log F_{5000} = -15.60$ at the top to $\log F_{5000} =
-16.74$ at the bottom.
\label{dimages}
}
\end{figure}

The procedures used to reduce the data, identify line emitters, and measure
their brightnesses were identical to those detailed in \citet{c02} and
\citet{ipn2}.  After de-biasing, flat-fielding, and aligning the data, our
narrow-band frames were co-added to create a master image that was clipped
of cosmic rays.   This frame was then compared to a deep $B$+$V$ continuum 
image provided by the MUSYC survey \citep{musyc} in two different ways.
First, the DAOFIND task within IRAF was run on the summed narrow-band and 
continuum image using a series of three convolution kernels, ranging from
one matching the image point-spread-function (PSF), to one $\sim 3$ times 
larger.  This created a source catalog of all objects in our field.   These
targets were then photometrically measured with DAOPHOT's PHOT routine, and 
sources with on-band minus continuum colors less than $-1.03$ in the AB system 
were flagged as possible emission-line sources (see Figure~\ref{cmd}).  At 
the same time, candidate LAEs were also identified by searching for positive 
residuals on a ``difference'' image made by subtracting a scaled version of 
the $B$+$V$ continuum image from the narrow-band frame.  In this case, the
DAOFIND algorithm was set to flag all objects brighter than four times the
local standard deviation of the background sky (see Figure~\ref{dimages}).
As pointed out by \citet{ipn2}, these two techniques complement each other,
since each detects objects that the other does not.  Specifically, 
$\lesssim 10\%$ of galaxies were missed by the color-magnitude method 
due to image blending and confusion, but found with the difference method.
Conversely, objects at the frame limit that were lost amidst the increased
noise of the difference frame, could still be identified via their on-band
minus off-band colors.  

Finally, because we intentionally biased our DAOFIND parameters to identify 
faint sources at the expense of false detections, each emission-line candidate 
was visually inspected on the narrow-band, $B+V$ continuum, and difference 
frames, as well as two frames made from subsamples of half the on-band 
exposures.  This last step excluded many false detections at the frame
limit, and left us with a sample of 259 candidate LAEs for analysis.

Once found, the equatorial positions of the candidate emission-line galaxies
were derived with respect to the reference stars of the USNO-A 2.0 astrometric 
catalog \citep{monet}.  The measured residuals of the plate solution
were $\sim 0\farcs 2$, a number slightly less than the $0\farcs 25$ external
error associated with the catalog.  Relative narrow-band magnitudes for 
the objects were derived by first measuring the sources with respect to
field stars using an aperture slightly greater than the frame PSF\null.
Since most of the galaxies detected in this survey are, at best, marginally 
resolved on our $1\arcsec$ images, this procedure was sufficiently 
accurate for our purposes.  We then obtained standard AB magnitudes by 
comparing large aperture photometry of the field stars to similar measurements 
of the spectrophotometric standards Feige~56 and Hiltner~600 \citep{stone} 
taken on three separate nights.  The dispersion in the photometric zero point 
computed from our standard star measurements was 0.03~mag.

\subsection{Derivation of Monochromatic Fluxes}
The fast optics of wide-field instruments, such as the MOSAIC camera
at the CTIO 4-m telescope, present an especially difficult challenge
for narrow-band imaging.  The transmission of an interference filter depends
critically on the angle at which it is illuminated:  light entering at the
normal will constructively/destructively interfere at a different
wavelength than light coming in at an angle \citep{eather}.  As a result, when
placed in a fast converging beam, an interference filter will have its bandpass
broadened and its peak transmission decreased by a substantial amount.
This effect is important, for without precise knowledge of the filter
bandpass, it is impossible to derive accurate monochromatic fluxes or
estimate equivalent widths.

\begin{figure}
\figurenum{4}
\plotone{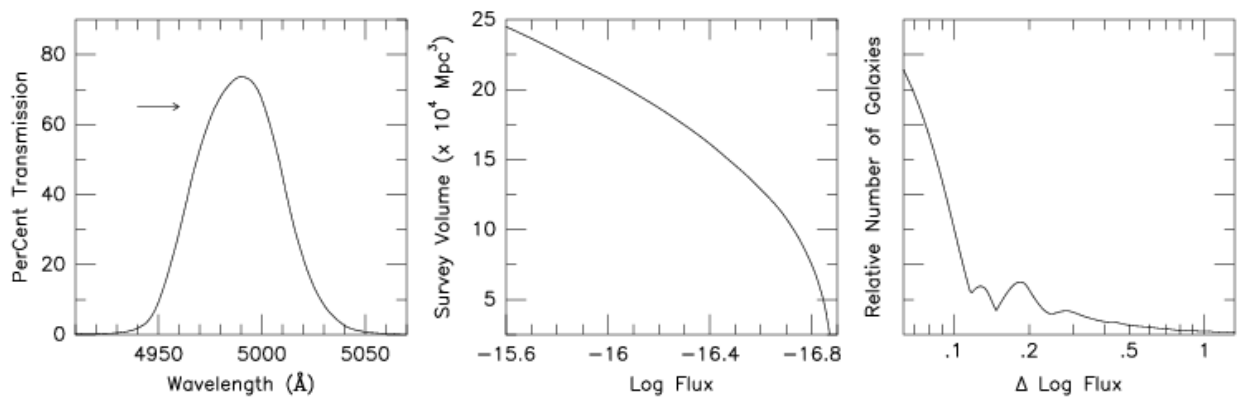}
\figcaption[filtresp]
{The left-hand panel shows the transmission curve for our
narrow-band $\lambda 5000$ filter at the outside ambient temperature
and in the converging f/3.2 beam of the 4-m telescope.  Note that the
bandpass is nearly Gaussian in shape.  The arrow shows the transmission
value used to translate AB magnitude into monochromatic flux
(see text).   The center panel uses the transmission function to illustrate
how our survey volume changes with emission-line sensitivity.  The
right-hand panel translates the transmission function into the photometric
convolution kernel that is described in the text.
\label{filtresp}
}
\end{figure}

To derive the filter transmission, we began with the throughput information
provided by the CTIO
observatory\footnote{http://www.ctio.noao.edu/instruments/FILTERS/index.html}. 
This curve, which represents the expected transmission of the 
[O~III] interference filter in the f/3.2 beam of the Blanco telescope, was 
computed by combining laboratory measurements of the filter tipped at several 
different angles from the incoming beam \citep[for a discussion of this 
procedure, see][]{m81}.  We then shifted this curve 2~\AA\ to the blue, to 
compensate for the thermal contraction of the glass at the telescope, and 
compared this model bandpass to the measured emission-line wavelengths obtained
from follow-up spectroscopy \citep{lira07}.   Interestingly, redshift 
measurements of 72~galaxies detected in three independent MUSYC fields confirm 
the shape of the filter's transmission curve, but not its central wavelength: 
according to the spectroscopy, the mean wavelength of the filter is 10~\AA\ 
bluer than given by CTIO \citep{gawiser07}.  Examining the source of this 
discrepancy is beyond the scope of this paper.  However, the data do confirm 
that, when placed in the beam of the CTIO 4-m prime focus MOSAIC camera, the 
bandpass of the CTIO [O~III] interference filter is nearly Gaussian in shape.
This bandpass is reproduced in the left-hand panel of Figure~\ref{filtresp}.

This non-square bandpass has important consequences for the analysis of
large samples of emission-line galaxies.  The first of these involves 
the definition of survey volume.    Because the transmission of the filter 
declines away from the bandpass center, the volume of space sampled by
our observations is a strong function of line strength.   This is illustrated
in the center panel of Figure~\ref{filtresp}.  Objects with bright line
emission can be detected even if their redshifts place Ly$\alpha$
in the wings of the filter, hence the volume covered for these objects is
realtively large.   Conversely, weak Ly$\alpha$ sources must have their line 
emission near the center of the bandpass to be observable.   As a result, the
``effective'' volume for our integrated sample of galaxies is a function of 
the galaxy emission-line luminosity function.

A second concern deals with the sample's flux calibration.
In order to compare the flux of an emission-line
object to that of a spectrophotometric standard star (\ie\ a continuum
source) one needs to know both the filter's integral transmission and its 
monochromatic transmission at the wavelength of interest \citep{jqa, m81}. 
When observing objects at known redshift, the latter requirement is not an 
issue.  However, when measuring a set of galaxies which can fall anywhere 
within a Gaussian-shaped transmission curve, the transformation between an 
objects' (bandpass-dependent) AB magnitude and its monochromatic flux is not 
unique.  In fact, if we assume that galaxies are (on average) distributed 
uniformly in redshift space, then the number of emission-line objects present 
at a given transmission, $T$, is simply proportional to the amount of 
wavelength associated with that transmission value.  Consequently, the observed 
distribution of emission-line fluxes will be related to the true 
distribution via a convolution, whose (unity normalized) kernel, $G(T)$, is
\begin{equation}
G(T) dT  = \left\{ \Bigg|{d \lambda \over dT} \Bigg| dT \right\}_{\rm
blue} +
\left\{ \Bigg|{d \lambda \over dT} \Bigg| dT \right\}_{\rm red}
\end{equation}
where the first term describes the filter's response blueward of the
transmission peak and the second term gives the response redward of
the peak.  The center panel of Figure~\ref{filtresp} displays this kernel
for the filter used in our survey.  The curve shows that for roughly half of
the detectable galaxies in our field, the effect of our filter's non-square 
bandpass is minimal.  However, for the other $\sim 50\%$ of galaxies, the 
shape of the bandpass is extremely important, and the inferred fluxes for some
objects can be off by over a magnitude.

Any analysis of the ensemble properties of our LAEs must consider the full
effect that the non-square bandpass and the odd-shaped convolution kernal
has on the sample.  We do this in Sections 4 and 5.  However, one often wants 
to quote the monochromatic flux and equivalent width for an {\it individual\/} 
Ly$\alpha$ emitter.  To do this, we need to adopt an appropriate ``mean'' value
for the transmission of our filter.  The most straightforward way
to define this number is via the filter's peak transmission.  This is 
where the survey depth is greatest, and choosing $T_{\rm max}$ is 
equivalent to assigning each galaxy its ``most probable'' monochromatic 
flux.   Unfortunately, by defining the transmission in this way, 
we underestimate the flux from all galaxies whose line emission does not 
fall exactly on this peak.  Alternatively, we can attempt to choose a
transmission which globally minimizes the flux errors of all the
galaxies detected in the survey.  This can be done by weighting each
transmission by the number of galaxies one expects to observe at that 
wavelength: the greater the transmission, the deeper the survey, and the more 
galaxies present in the sample.  The difficulty with this ``expectation
value'' approach is that it requires prior knowledge of the distribution of 
emission-line fluxes, which is one of the quantities we are attempting 
to measure.  That leads us to a third possibility:  to approximate the
filter's expectation value using some ``characteristic'' transmission, $T_C$, 
which is independent of the galaxy luminosity function, but still takes
the filter's changing transmission into account.   The arrow in 
Figure~\ref{filtresp} identifies the transmission we selected as being 
characteristic of the filter; the justification for this value is presented
in Section~5.  We emphasize that $T_C$ is only a convenient mean that
enables us to quote the likely emission-line strengths of {\it individual\/} 
galaxies.  When analyzing the global properties of an ensemble of LAEs, the
full non-Gaussian nature of the filter's convolution kernel must be
taken into account. 

Using this transmission and our knowledge of the filter curve, we
converted the galaxies' AB magnitudes to monochromatic fluxes at 
$\lambda = 5000$~\AA\ via
\begin{equation}
F_{5000} = 3.63 \times 10^{-20} \, 10^{-m_{\rm AB} / 2.5} \, \cdot \,
{c \over \lambda^2} \, \cdot \, {\int T_{\lambda} d\lambda \over T_C}
\end{equation}
where $F_{5000}$ is given in ergs~cm$^{-2}$~s$^{-1}$ \citep{jqa}.  
Equivalent widths then followed via
\begin{equation}
{\rm EW} = {F_{5000} \over f_{B+V}} - \Delta \lambda
\end{equation}
where $f_{B+V}$ is the objects' AB flux density in the $B+V$ continuum image, 
and $\Delta\lambda$, the FWHM of the narrow-band filter, represents the
contribution of the galaxy's underlying continuum within the bandpass.   Both 
these equations are only applicable to objects whose line emission dominates 
the continuum within the narrow-band filter's bandpass.   Since we are limiting 
our discussion to galaxies with narrow-band minus broad-band AB magnitudes 
more negative than $-1.03$, this approximation is certainly valid.   However, 
we do note that by using $T_C$ instead of $T_{\rm max}$, we are intentionally 
overestimating the flux and equivalent width of some galaxies, in order to 
minimize the errors in others.   So, while the application of $T_C$ formally
translates our $\Delta m = -1.03$ criterion into a minimum emission-line
equivalent width of 90~\AA, galaxies with emission-lines that fall near the 
peak of the filter transmission function can have equivalent widths that are
$\sim 12\%$ smaller.   This implies that the absolute minimum equivalent 
width limit for our sample of LAEs is 80~\AA.

\subsection{Sample of LAE Candidates}

Tables~\ref{brightLAEs} and \ref{faintLAEs}
give the coordinates of each candidate emission-line
galaxy, along with its inferred monochromatic flux and equivalent width.  
In total, 259 objects are listed, though many are beyond the limit of
our completeness.  To determine this limit, we followed the procedures of 
\citet{ipn2} and added 1,000,000 artificial stars (2000 at a time) to our 
narrow-band frame.  By re-running our detection algorithms on these modified 
frames, we were able to compute the flux level below which the object recovery 
fraction dropped below the 90\% threshold.   This value, which corresponds
to a monochromatic flux of $1.5 \times 10^{-17}$~ergs~cm$^{-2}$~s$^{-1}$
($\log F_{5000} = -16.82$) is our limiting magnitude for statistical
completeness; 162 galaxies satisfy this criterion.  

Before proceeding further with our analysis, we performed one additional check
on our data.  To eliminate obvious AGN from our sample, we cross-correlated
our catalog of emission-line objects with the lists of X-ray sources found
in the 1~Msec exposure of the Chandra Deep Field South \citep{alexander},
and the four 250~ksec exposure of the Extended Chandra Deep Field South
\citep{lehmer05, virani}.  Two of our LAE candidates were detected in the X-ray
band.  The first, which is our brightest Ly$\alpha$ emitter, has a 0.5 -- 8 keV
flux of $3.4 \times 10^{-15}$ ergs~cm$^{-2}$~s$^{-1}$ (\ie\ $L_{\rm X} \sim
2.8 \times 10^{44} \, \h7^{-2}$~ergs~s$^{-1}$ at $z = 3.1$) and exhibits
C~IV emission at 1550~\AA\ \citep{lira07}.  The other is an interloper:
a $z = 1.6$~AGN detected via its strong C~III] line at 1909~\AA\null.  
For the remaining 160 objects that were not detected individually in the X-ray
band, we used stacking analyses to constrain their mean X-ray power output
\citep[see][for details]{lehmer07}.  We find that the stacked X-ray
signal, which corresponds to a $\sim 40$~ksec effective exposure on an
average LAE, does not yield a $3 \sigma$ detection in any of three
X-ray bandpasses (\hbox{0.5--8.0~keV}, \hbox{0.5--2.0~keV}, and 
\hbox{2--8~keV}).  These results imply a $3 \sigma$ upper-limit of
$\sim 3.8 \times 10^{41}$~$h_{70}^{-2}$~ergs~s$^{-1}$ on the mean
\hbox{0.5--2.0~keV} luminosity for our LAEs, which demonstrates that few
of our Ly$\alpha$ sources harbor low-luminosity AGN\null.  Similarly, 
if we use the conversion of \citet{ranalli}, we can translate this
X-ray non-detection into an upper-limit for a typical LAE's 
star-formation rate.  This limit, $85 \, \h7^{-2} \, M_{\odot}$~yr$^{-1}$, 
is roughly an order of magnitude greater than the rates inferred from the
objects' Ly$\alpha$ emission or UV continua (see Section~6).

For the remainder of this paper, we will treat our $z = 3.1$ X-ray source
as AGN and exclude it from the analysis.  This leaves us with a sample of 
160 objects, which we assume are all star-forming galaxies.  We note that,
because all of our objects have equivalent widths greater than
80~\AA, they are unlikely to be [O~II] emitters.  At $z \sim 0.34$,
our survey volume is only $\sim 7300 \, \h7^{-3}$~Mpc$^{3}$,
which, through the luminosity functions of \citet{hogg98}, \citet
{gallego02}, and \citet{teplitz03}, implies a total population of 
between $\sim 20$ and $\sim 200$ [O~II] emission-line galaxies above 
our completeness limit.  Since less than 2\% of these objects will have 
rest frame equivalent widths greater than $\sim 60$~\AA\ \citep{hogg98}, 
the number of [O~II] interlopers in our sample should be negligible.  
This estimate is confirmed by follow-up spectroscopy:  of the 52 LAE
candidates observed with sufficient signal-to-noise for a redshift
determination, {\it all\/} are confirmed Ly$\alpha$ emitters
\citep{gawiser, lira07}.

\begin{figure}
\figurenum{5}
\plotone{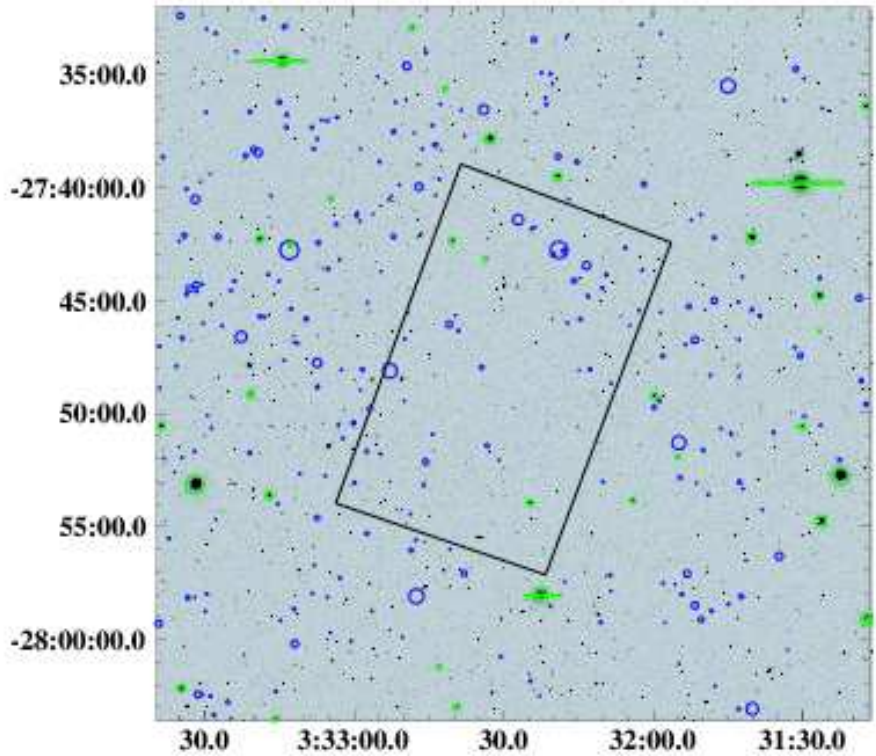}
\figcaption[map]
{The sky coordinates of the 160~candidate $z = 3.1$ LAEs
brighter than our completeness limit plotted over our narrow-band
5000~\AA\ image.  The size of each circle is proportional
to Ly$\alpha$ luminosity, with the largest circle representing
$1.25 \times 10^{43} \, \h7^{-2}$~ergs~s$^{-1}$.  The green regions show 
areas of the chip near bright stars that were excluded from the analysis; 
the large rectangle is the GOODS field.
\label{map}}
\end{figure}

Figure~\ref{map} shows the spatial distribution of the LAEs
above our completeness limit.  The sources are obviously clustered,
falling along what appear to be ``walls'' or ``filaments''.  The
GOODS region has a below-average number of $z = 3.1$ Ly$\alpha$ emitters, 
and there are almost no objects in the northwestern part of the field.  
Conversely, the density of LAEs east and northeast of the field center is 
quite high.  This type of data can be an extremely powerful probe of 
cosmological history, but we will defer a discussion of this topic
to a future paper \citep{gawiser07}.

\section{The Continuum Properties of the Emitters}

\begin{figure}
\figurenum{6}
\plotone{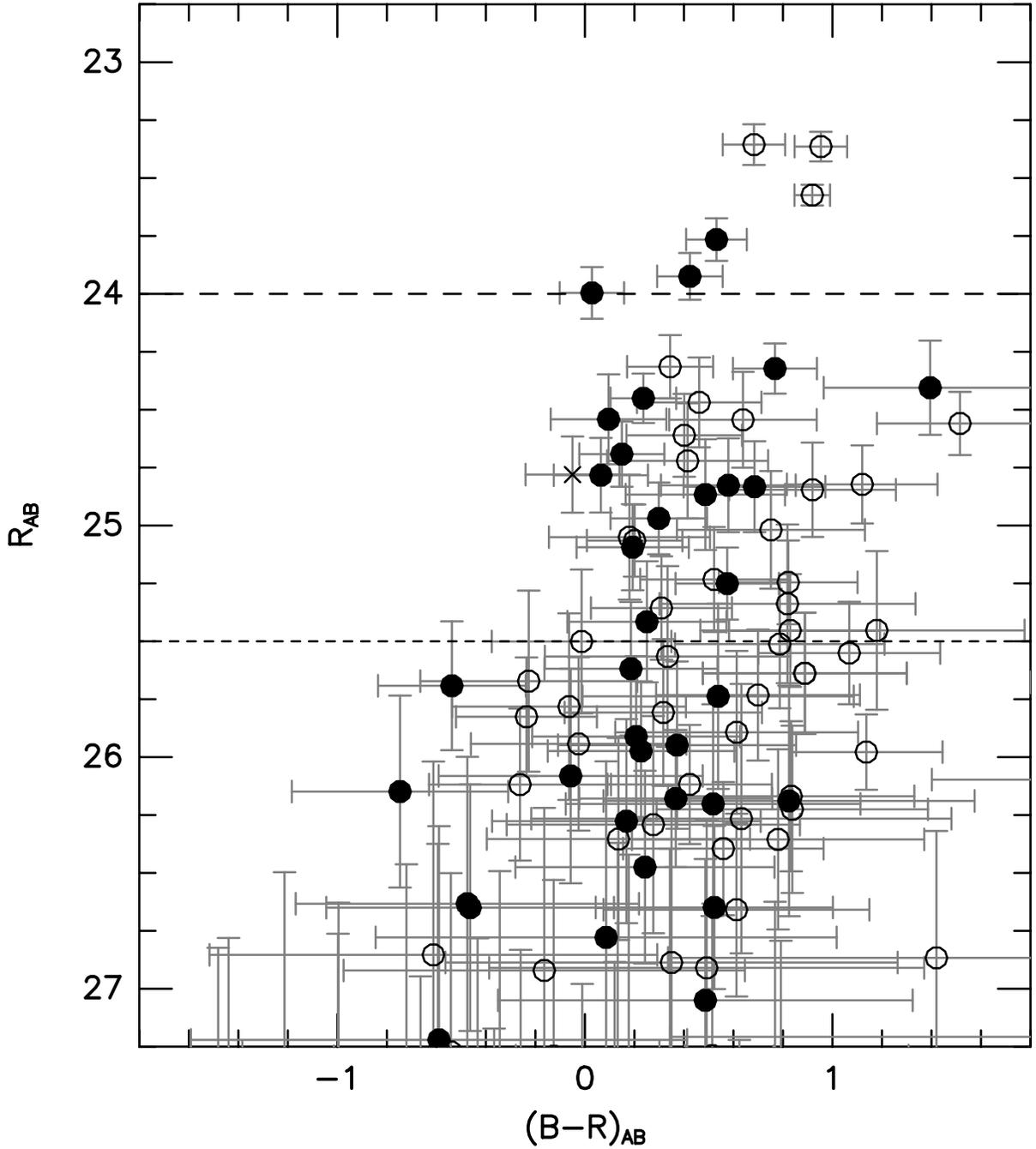}
\figcaption[bmr]
{The $B-R$ (rest frame $m_{1060} - m_{1570}$) color-magnitude diagram for
LAEs.  The solid circles represent galaxies which have been spectroscopically
confirmed as Ly$\alpha$ emitters \citep{lira07}; the cross indicates the lone
AGN\null.  The long-dashed line at $R = 24$ represents the typical magnitude 
limit of LBG spectroscopic surveys; the short-dashed line at $R = 25.5$ gives 
the photometric limit of most LBG observations.  Note that the median
color of our LAEs is quite blue; this is consistent with models 
for galaxies with recent star formation.  Note also
the large range of colors displayed in the figure.  This scatter is 
greater than that expected from the photometric errors, and suggests that 
the LAE population is not homogeneous.
\label{bmr}}
\end{figure}

To investigate the continuum properties of our Ly$\alpha$ emitters, we
measured the brightness of each LAE on the broadband UBVR images of the
MUSYC survey \citep{musyc}.  Since the catalog associated with this dataset
has a $5\,\sigma$ detection threshold of $U = 26.0$, $B = 26.9$, $V = 26.4$, 
and $R = 26.4$, our knowledge of the LAEs' positions (obtained from
the narrow-band frames) allows us to perform photometry well past this limit.
Figure~\ref{bmr} displays the $B-R$ color-magnitude diagram for 88 of the LAEs 
brighter than $R_{AB} = 27.25$.  The diagram, which shows the 
galaxies' rest-frame continua at 1060 and 1570~\AA, has several features 
of note.

The first involves the color distribution of our objects.  According to 
the figure, LAEs with $R$-band magnitudes brighter than $R = 25$ have a 
median color of $B-R = 0.53$.  This value agrees with the blue colors found 
by \citet{venemans} for a sample of Ly$\alpha$ emitters at $z = 3.13$, and
is the value expected for a $\sim 10^8$~yr old stellar system evolving with 
a constant star-formation rate \citep{fujita, bc03}.   This median color
is also consistent with the results of \citet{gawiser}, who stacked the 
broad-band fluxes  of 18 spectroscopically confirmed $z = 3.1$ LAEs and
showed that the typical age of these systems is between $0.01 < t < 2$~Gyr.  
It does, however, stand in marked contrast to the results of 
\citet{stiavelli}, who claimed that Ly$\alpha$ emitters at $z = 2.4$ are 
very red ($B-I \sim 1.8$).  The blue colors of our galaxies confirm their 
nature as young, star-forming systems.  There is no evidence for excessive 
reddening in these objects, and if the galaxies do possess an underlying 
population of older stars, the component must be quite small.

On the other hand, as the LAE color distribution indicates, Ly$\alpha$
emitters are not, as a class, homogeneous.  At $R = 25$, the MUSYC $B-R$ 
colors have a typical photometric uncertainty of $\sigma_{B-R} = 0.25$~mag. 
This contrasts with the observed color dispersion for our galaxies, which is
$\sim 0.4$~mag for objects with $R < 25$.  Thus, there is at least a 
$\sim 0.3$~mag scatter in the intrinsic colors of these objects.  Either there 
is some variation in the star-formation history of Ly$\alpha$ emitters, or
dust is having an effect on the emergent colors.

\begin{figure}
\figurenum{7}
\plotone{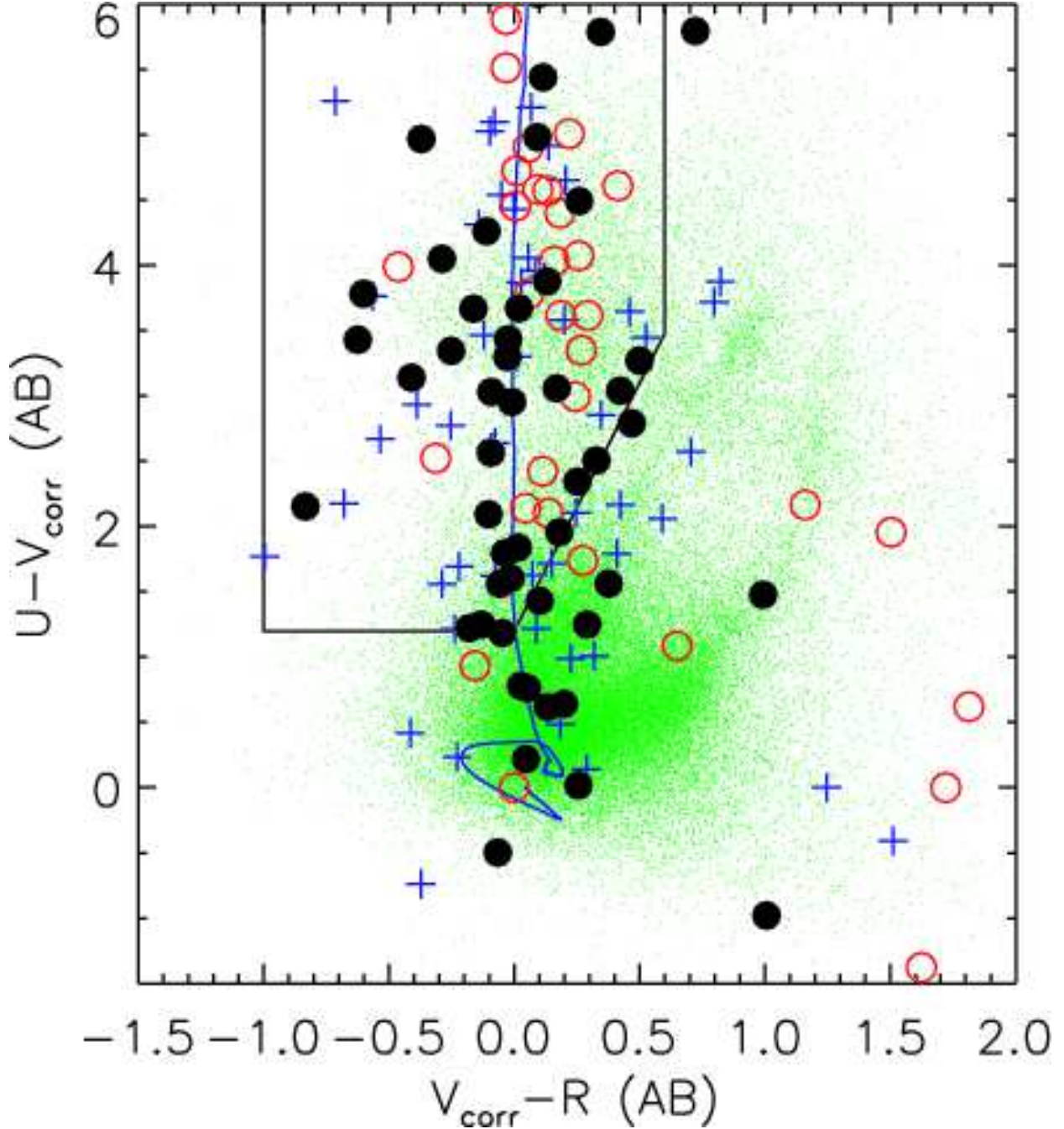}
\figcaption[lbgcomp]
{The $U-V$ versus $V-R$ colors of our $z = 3.1$ LAEs.  The solid circles
show spectroscopically confirmed LAEs, the open circles represent
sources observed with insufficient signal-to-noise for classification,
and the crosses are objects with no spectroscopy.  The dots are the
entire 84,410 object catalog.  The polygon is the LBG
selection region; the sold curve is the track of an LBG template
spectrum.  This track falls inside the selection region in the redshift
range $2.8 < z <3.4$ \citep{shapley}.   Although most LAEs have LBG-like 
colors, their $R > 25.5$ magnitudes exclude them from the ``spectroscopic''
samples studied by \citet{steidel96a, steidel96b, steidel03}.
The contribution of each LAE's  emission line to its
$V$-band flux has been subtracted to yield $V_{corr}$.
\label{lbgcomp}}
\end{figure}

Finally, it is worth emphasizing that our Ly$\alpha$ emitters are substantially
fainter in the continuum than objects found by the Lyman-break technique.
At $z \sim 3$, $L^*$ galaxies have an apparent magnitude of 
$R \sim 24.5$ \citep{steidel99} and ground-based Lyman-break surveys 
typically extend only $\sim 1$~mag beyond this value
\citep[see][for a review]{giavalisco}.  Furthermore, spectroscopic surveys of
LBG candidates rarely target galaxies fainter than $R = 24$.  In our 
emission-line sample, the median continuum magnitude is $R \sim 26.7$, and 
many of the galaxies have aperture magnitudes significantly fainter than 
$R \sim 28$.   In general, LAEs do inhabit the same location as LBGs in the
$U$-$V$ vs.~$V$-$R$ color-color space (see Figure~\ref{lbgcomp}), but
their extremely faint continuum sets them apart.

\begin{figure}
\figurenum{8}
\plotone{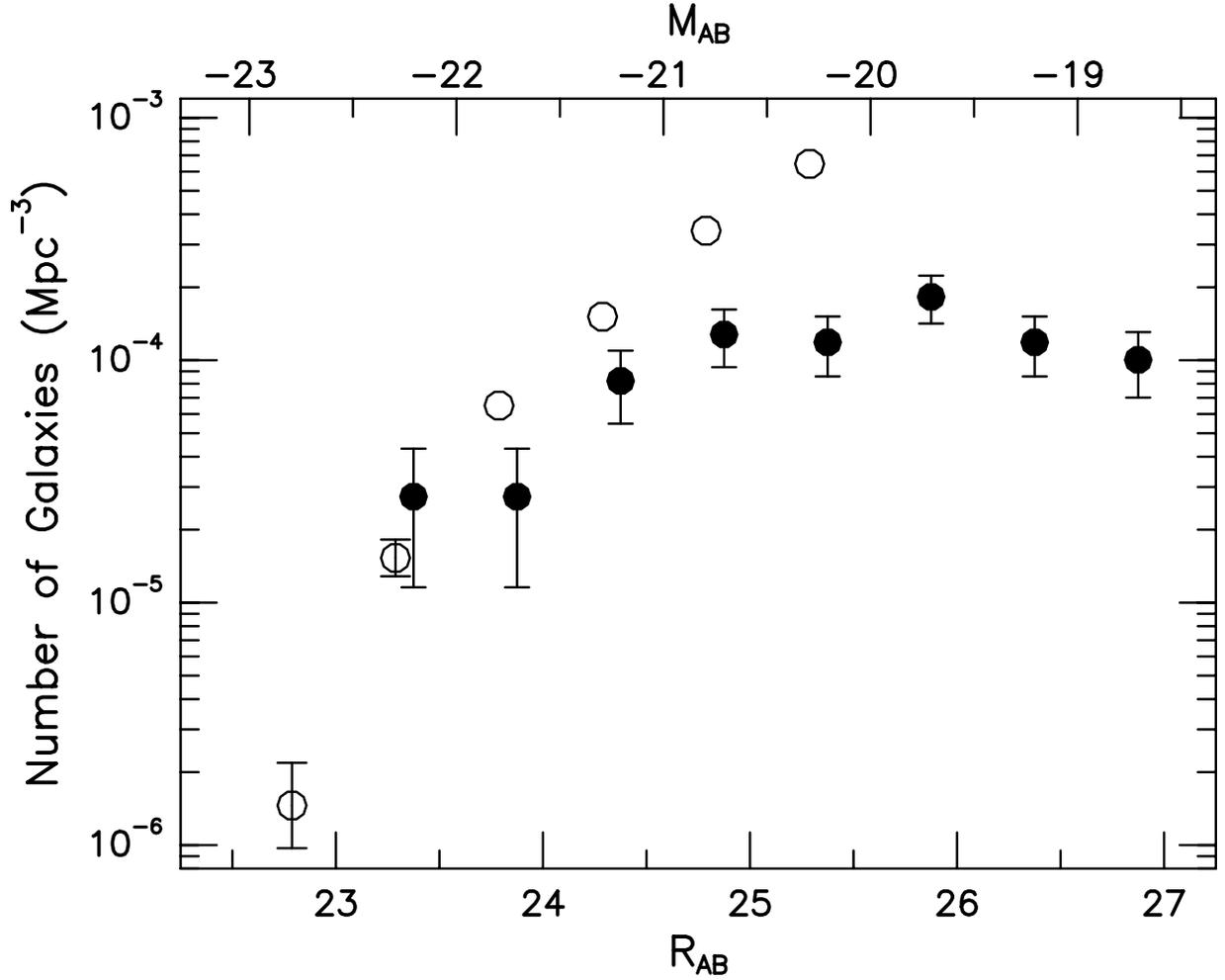}
\figcaption[continuum]
{The $R_{AB}$ (rest frame 1570~\AA) luminosity function of our $z = 3.1$
Ly$\alpha$ emitters (solid circles), compared to the rest-frame 1700~\AA\ 
luminosity function of $z = 3.04$ Lyman-break galaxies (open circles) from
\citep{steidel99}.  
The flattening of our luminosity function at $R > 26.5$ is due to selection:
at these magnitudes, only the strongest line emitters make it into
our sample.  In the magnitude range $R < 25.5$, $z = 3.1$ Ly$\alpha$ 
emitters are $\sim 3$ times rarer than Lyman-break galaxies.
\label{continuum}}
\end{figure}

This is also illustrated in Figure~\ref{continuum}, which compares the
rest frame 1570~\AA\  luminosity function of our complete sample of
Ly$\alpha$ emitters (those with monochromatic fluxes greater than
$1.5 \times 10^{-17}$~ergs~cm$^{-2}$~s$^{-1}$) with the rest-frame
1700~\AA\ luminosity function of $z = 3.1$ Lyman-break galaxies
\citep{steidel99}.  When plotted in this way, our sample of LAEs appears
incomplete, since for $R \gtrsim 26.5$, only the brightest emission-line
sources will make it into our catalog.   The plot also implies that at
$z = 3.1$, $R < 25.5$, Ly$\alpha$ emitters are $\sim 3$ times rarer than
comparably bright Lyman-break galaxies.  Since this ratio is virtually 
identical to that measured by \citet{steidel00} within an extremely rich
$z = 3.09$ protocluster, this suggests that the number is not a strong
function of galactic environment.  But, most strikingly, our observations
demonstrate the Ly$\alpha$ emitters sample the entire range of the 
(UV-continuum) luminosity function.   The median UV luminosity of LAEs in our
sample is $\lesssim 0.2 L^*$, and the faintest galaxy in the group is no
brighter than $\sim 0.02 L^*$.  Just as broadband observations 
detect all objects at the bright-end of the continuum luminosity function, but
sample the entire range of emission-line strengths, our narrow-band
survey finds all the brightest emission-line objects, but draws from the 
entire range of continuum brightness.

\section{The Ly$\alpha$ Equivalent Width Distribution}

Before examining the emission-line properties of our dataset,
we need to correct for the observational biases and selection 
effects that are present in the sample.  Since the data were taken
in a fast-beam through a filter with a non-square bandpass, these effects are 
substantial.  Continuum measurements, of course, are unaffected by the 
peculiarities of a narrow-band filter, but the distribution of monochromatic 
fluxes can be significantly distorted.  Specifically, the observed flux
distribution will be  the convolution of the true distribution with the 
following two kernels:

{\it The Photometric Error Function: } The random errors associated with
our narrow-band photometry vary considerably, ranging from
$\sim 0.02$~mag at the bright end, to $\sim 0.2$~mag near the completeness
limit (see Table~\ref{errors}).  
These errors will scatter objects from heavily 
populated magnitude bins into bins with fewer objects, and flatten the
slope of the luminosity function.  Because the change in slope goes as the 
square of the measurement uncertainty \citep{eddington13, eddington40}, the 
effect of this convolution is most important for objects near the survey
limit.

{\it The Filter Transmission Function.}  As described in Section 2.1, the
narrow-band filter used for this survey has a transmission function that is 
nearly Gaussian in shape.  This creates an odd-shaped convolution kernel (the
right panel of Figure~\ref{filtresp}), which systematically decreases the 
measured line-emission of objects falling away from the peak of the 
transmission curve.  Moreover, because the objects' equivalent widths are 
also reduced by this bandpass effect, some fraction of the LAE population will 
be lost from our EW $> 80$~\AA\ sample.  The result is that the 
normalization of this filter transmission kernel is not unity.  Instead, it
depends on the intrinsic equivalent width distribution of the galaxies, since 
that is the function that defines the fraction of galaxies (at each redshift) 
which can still make it into our sample.

These effects are illustrated in the top panel of Figure~\ref{ewhist}, which 
displays a histogram of the rest-frame equivalent widths for our candidate 
Ly$\alpha$ galaxies.  As the dotted line shows, the data appear to be well 
fit by an exponential that has an e-folding length of 
$w_{\rm obs} = 214^{+19}_{-15}$~\AA\null.  However, because
the bandpass of our narrow-band filter is more Gaussian-shaped than 
square, the line-strengths of many of the galaxies have been 
systematically underestimated.  In fact, the true distribution of equivalent 
widths is broader than that measured:  when we perform a 
maximum-likelihood analysis using a series of exponential laws, convolved 
with the filter bandpass and photometric error kernels, we obtain a
most-likely scale length of $w_{\rm obs} = 311^{+47}_{-33}$~\AA, or
$w_0 = 76^{+11}_{-8}$~\AA\ in the rest frame of the sample.

Such a distribution is quite different from that reported by 
\citet{mr02}.   In their survey of 150 $z = 4.5$ Ly$\alpha$ emitters,
$\sim 60\%$ of the objects had extremely high rest-frame equivalent widths,
\ie\ EW$_0 > 240$~\AA\null.  Since stellar population models, such as those
by \citet{cf93} cannot produce such strong line-emission, \citet{mr02}
postulated the presence of a top-heavy initial mass function and perhaps
the existence of Population~III stars.  However in our sample, only 3 
out of 160 LAEs ($\sim 2\%$) have observed rest-frame equivalent widths
greater than this 240~\AA\ limit.   Even when we correct for the effects
of our filter's non-square bandpass, the fraction of strong line-emitters does 
not exceed $\sim 12\%$.   This is less than the $\sim 20\%$
value estimated by \citet{dawson} via Keck spectroscopy of a subset 
of \citet{mr02} objects.  Thus, at least at $z \sim 3.1$, there is no 
need to invoke a skewed initial mass function to explain the majority of our
LAEs.

The equivalent width distribution of Figure~\ref{ewhist} also differs
dramatically from that found by \citet{shapley} for a sample of
$z \sim 3$ Lyman-break galaxies.  In their dataset, rest-frame equivalent
widths e-fold with a scale-length of $\sim 25$~\AA, rather than the
$\sim 75$~\AA\ value derived from our LAE survey.   This difference is
not surprising given that the former dataset is selected to be bright
in the continuum, while the latter is chosen to be strong in the
emission-line.  Moreover, when \citet{shapley} analyzed the $\sim 25\%$ 
of Lyman-break galaxies with rest-frame equivalent widths greater than 
20~\AA, they found a correlation between line strength and continuum 
($R$-band) magnitude, in the sense that fainter galaxies had higher 
equivalent widths.   We see that same trend in our data, but it is 
largely the result of a selection effect.  (Faint galaxies with low 
equivalent widths fall below our monochromatic flux limit.)  A comparison 
of emission-line flux with equivalent width for our statistically complete 
sample shows no such correlation. 

The lower two panels of Figure~\ref{ewhist} demonstrate this another way.
In the diagram, our sample of LAEs is divided in half, with the middle panel
showing the equivalent width distribution for objects with monochromatic
Ly$\alpha$ luminosities greater $2 \times 10^{42} \, \h7^{-2}$~ergs~s$^{-1}$, 
and the bottom panel displaying the same distribution for less luminous 
objects.  As the figure illustrates, the distribution of equivalent widths 
is relatively insensitive to the absolute brightness of the galaxy.  To 
first order this is expected, since both the UV continuum and the Ly$\alpha$ 
emission-line flux are driven by star formation.  However, one could imagine 
a scenario wherein the amount, composition, and/or distribution of dust within 
the brighter (presumably more-metal rich) Ly$\alpha$ emitters differs from 
that within their lower-luminosity counterparts.   Since the effect of this 
dust on resonantly-scattered Ly$\alpha$ photons is likely to be different
from that on continuum photons, this change in extinction can theoretically 
produce a systematic shift in the distribution of Ly$\alpha$ equivalent 
widths.   There is no evidence for such a shift in our data; this constancy 
argues against the importance of dust in these objects.

\section{The Ly$\alpha$ Emission-Line Luminosity Function}

\begin{figure}
\figurenum{10}
\plotone{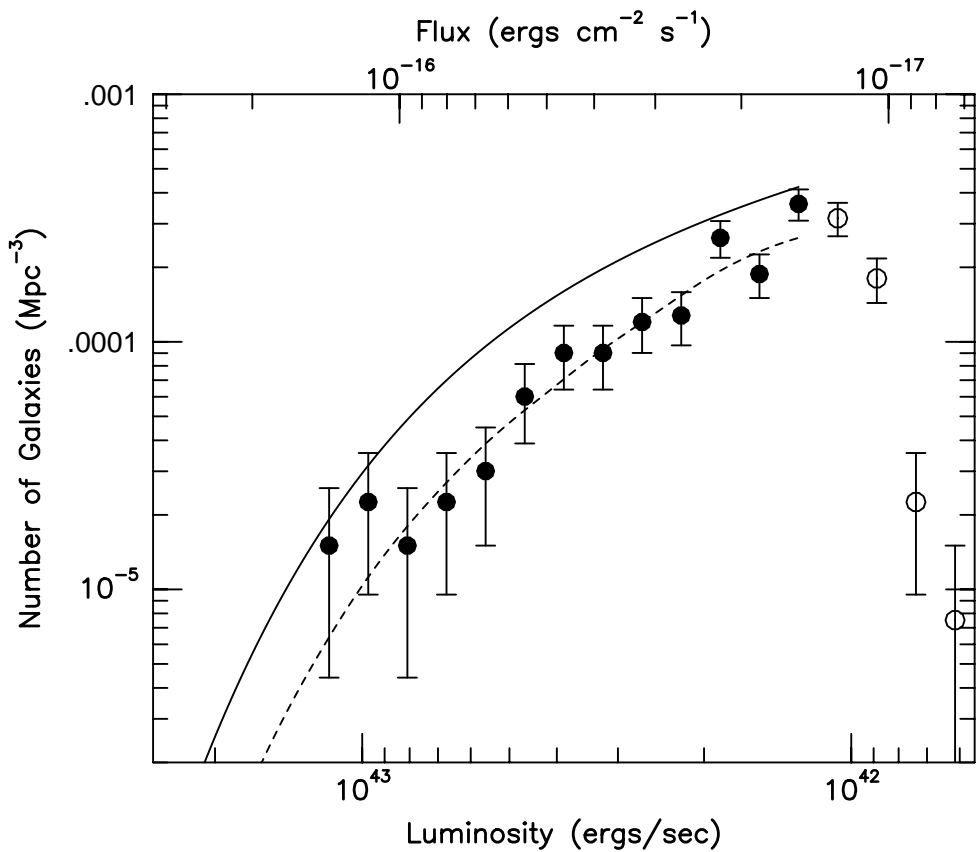}
\figcaption[lumfun]
{The number density of $z=3.1$ Ly$\alpha$ galaxies with observers' frame
equivalent widths greater than 90~\AA\ binned into 0.2~mag intervals.  The
points give the density of objects under the assumption that our filter's
FWHM defines the survey volume; the open circles represent data beyond
our completeness limit.  The solid curve shows our input best-fit 
\citet{schechter} luminosity function, while the dashed line illustrates 
the shape and normalization of this function after correcting for the effects 
of photometric error and our filter's non-square transmission curve.
\label{lumfun}}
\end{figure}

Figure~\ref{lumfun} shows the distribution of monochromatic fluxes for our
sample of emission-line galaxies.  The function looks much like a power
law, with a faint-end slope of $\alpha \sim -1.5$ that steepens as one moves
to brighter luminosities.  However, to quantify this behavior, we once again
have to correct the observed flux distribution for the distortions caused by 
photometric errors and the non-square bandpass of the filter.  In addition, we 
must also consider the censoring effect our equivalent width cutoff has on the 
data:  some line emitters whose redshifts are not at the peak of the filter 
transmission function will fall out of our sample completely.  

To deal with these effects, we fit the observed distribution of Ly$\alpha$
emission-line fluxes to a \citet{schechter} function via the method
of maximum likelihood \citep[\eg][]{hw87, p2}.  We applied our two 
convolution kernels (including the equivalent width censorship) to a series 
of functions of the form
\begin{equation}
\phi(L) d(L/L^*) \propto \left(L/L^*\right)^\alpha e^{-L/L^*} d(L/L^*)
\end{equation}
treated each curve as a probability distribution (\ie\ with a unity
normalization), and computed the likelihood that the observed sample of
Ly$\alpha$ fluxes is drawn from the resultant distribution.  The results for
the three parameters of this fit, $\alpha$, $\log L^*$, and $N$, the integral
of the Schechter function down to our limiting flux (in units of
galaxies~Mpc$^{-3}$), are shown in Figure~\ref{contours};
Table~\ref{parameters} lists the best-fitting parameters, along with their
marginalized most-likely values and uncertainties.  For completeness,
Table~\ref{parameters} also gives the value of $\phi^*$ which is inferred
from our most likely solution.  As expected, the plots 
illustrate the familiar degeneracy between $L^*$ and $\alpha$:  our best-fit 
solution has $\alpha \sim -1.5$, but if $L^*$ is forced to brighter 
luminosities, $\alpha$ decreases.  The contours also demonstrate an asymmetry 
in the solutions, whereby extremely bright values of $L^*$ are included within 
the $3 \, \sigma$ contours of probability, but faint values of the same 
quantity are not. 

But perhaps the most interesting feature of the analysis concerns the effective
volume of our survey.   As in Section 2.1, the amount of space sampled by the
observations depends critically on each galaxy's Ly$\alpha$ luminosity and
equivalent width.  Bright line-emitters with large equivalent widths
can be identified well onto the wings of the filter, hence the survey 
volume associated with these objects is relatively large.  Conversely,
weak line-emitters, and objects with small equivalent widths can only
be detected if they lie at the peak of the filter transmission curve.
Thus, the survey volume for these objects is quite small.  The effective 
volume for our observations is therefore a weighted average, which 
depends on the intrinsic properties of entire LAE sample.

This average can be computed from the data displayed in Figure~\ref{contours}.
According to the figure, the space density of galaxies with emission-line
brighter than $1.5 \times 10^{-17}$~ergs~cm$^{-2}$~s$^{-1}$  (\ie\ $1.3 
\times 10^{42} \, \h7^{-2}$~ergs~s$^{-1}$) is extremely well-defined, 
$1.46 \pm 0.12 \times 10^{-3} \, \h7^{3}$~galaxies~Mpc$^{-3}$.  Since this
measurement comes from the detection of 160 galaxies brighter than the
completeness limit, the data imply an effective survey volume of 
$\sim 1.1 \times 10^{5}~\h7^{-3}$~Mpc$^{3}$.  This is {\it not\/}
the volume one would infer from the interference filter's full-width
at half-maximum:  it is 25\% smaller, or roughly the full-width of 
the filter at two-thirds maximum.  

This difference is illustrated in Figure~\ref{lumfun}.  The points
show the space density of Ly$\alpha$ galaxies one would derive simply
by using the filter's FWHM to define the survey volume; the solid line
gives the \citet{schechter} function which best fits the data.  The
offset between the solid line and the dashed line, which represents the
function after the application of the two convolution kernels, confirms the
need for careful analysis when working with narrow-band data taken through a 
non-square bandpass.

The results of our maximum-likelihood calculation also suggest a simple 
definition for the effective transmission for our filter.  As described in 
Section~2.1, a ``characteristic'' transmission is needed to convert the
(bandpass-dependent) AB magnitude of an individual galaxy to monochromatic 
Ly$\alpha$ flux.  Rather than use the maximum transmission (which would 
underestimate the flux of all galaxies not at the filter peak), or adopt some 
complicated scheme which involves iterating on the luminosity function, one 
can simply choose the filter's mean transmission within some limited 
wavelength range.  Based on the results above, the filter's full-width at 
two-thirds maximum seems an appropriate limit.  This transmission, which is 
indicated by the arrow in Figure~\ref{filtresp}, is the value used to derive 
the fluxes and equivalent widths of Tables~\ref{brightLAEs} and 
\ref{faintLAEs}.  If were to use to filter's peak transmission instead of
this characteristic value, the tabulated emission-line fluxes and equivalent
widths would all be $\sim 12\%$ smaller.

The error bars quoted above for the space density of Ly$\alpha$ emitters
represent only the statistical uncertainty of the fits.  They do not 
include the possible effects of large-scale structure within our survey
volume.  Specifically, if the linear bias factor for LAEs is two
\citep[see][for an analysis of the objects' clustering]{gawiser07} then 
the expected fluctuation in the density of Ly$\alpha$ emitters measured within 
a $\sim 10^5 \, \h7^{-3}$~Mpc$^{3}$ volume of space is $\sim 30\%$.  This
value should be combined in quadrature with our formal statistical 
uncertainty.

\begin{figure}
\figurenum{12}
\plotone{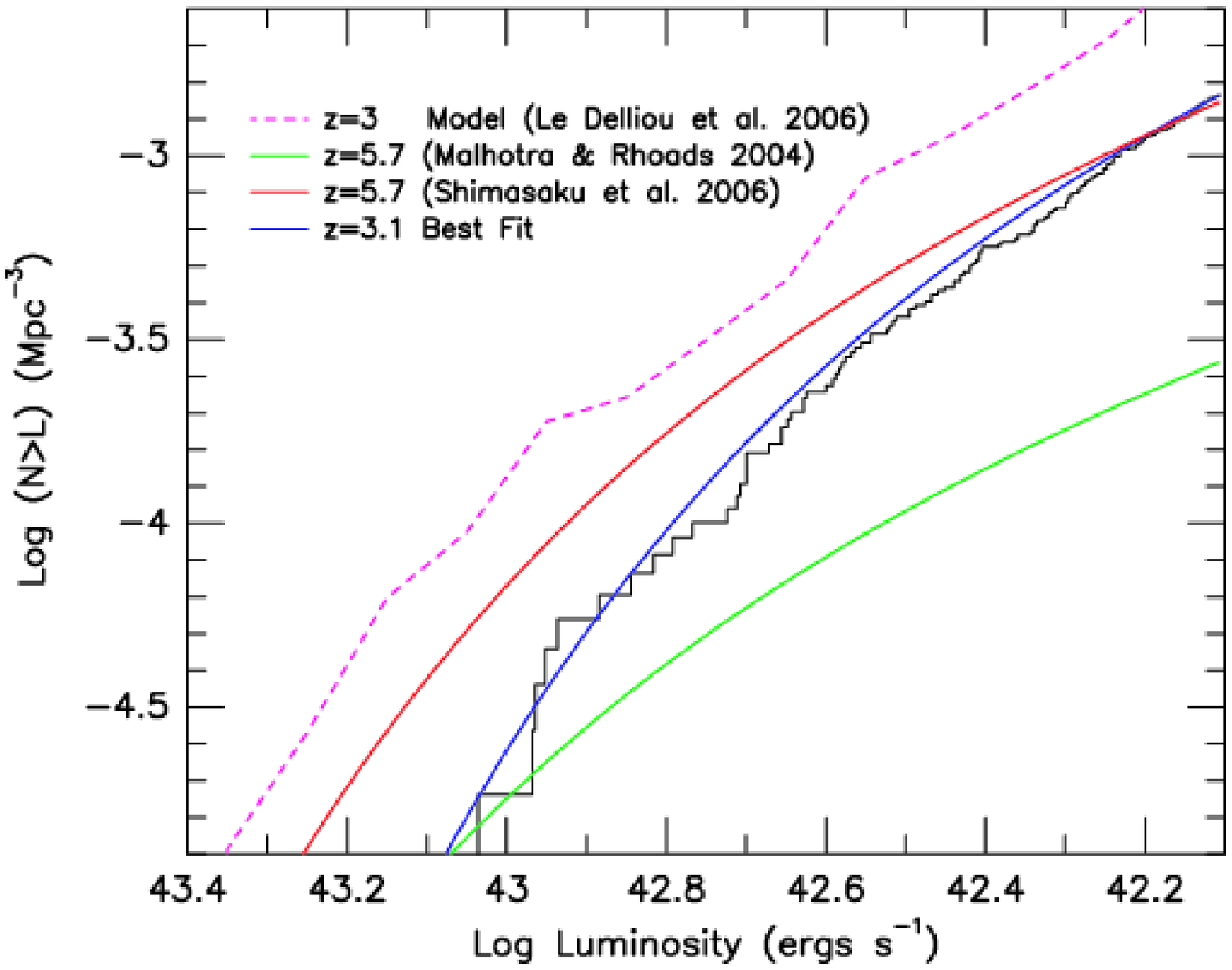}
\figcaption[cumfun]
{The cumulative Ly$\alpha$ luminosity function inferred from our
survey of Ly$\alpha$ emitters with rest-frame equivalent widths 
greater than 22~\AA.  The solid blue line shows our best fit 
\citet{schechter} function ($\alpha = -1.49$), the green line is the 
$\alpha = -1.5$ luminosity function found by \citet{mr04} for LAEs 
at $z = 5.7$, and the red line is the Schechter fit for $z = 5.7$
emitters found by \citet{shimasaku06}.   The dashed line is Model~A
by \citet{ledelliou06}.  For purposes of this figure, our data have been 
artificially normalized to match our best-fit function.   The large
difference between the \citet{mr04} and \citet{shimasaku06} fits makes it
impossible to study the evolution of the LAE population at this time.
\label{cumfun}}
\end{figure}

Since Ly$\alpha$ galaxies have been observed at a number of redshifts,
it is tempting to use our data to examine the evolution of the LAE 
luminosity function.  Unfortunately, the samples obtained to date are not 
yet robust enough for this purpose.  An example of the problem is shown in 
Figure~\ref{cumfun}, which compares our cumulative luminosity function 
(and our Schechter fit for $\alpha = -1.5$) to two measures of Ly$\alpha$ 
galaxies at $z = 5.7$.  As the figure illustrates, there are large differences
between the measurements.  If the \citet{mr04} luminosity function is
correct, then LAEs at $z = 3.1$ are a factor of $\sim 2.5$ brighter and/or 
more numerous than their $z = 5.7$ counterparts.   However, if the $z=5.7$ 
LAE luminosity function of \citet{shimasaku06} is correct, then evolution
is occurring in the opposite direction, \ie\ the star-formation rate density
is declining with time.   Without better data, it is difficult to derive any 
conclusions about the evolution of these objects.

Figure~\ref{cumfun} also plots our data against the predictions of a
hierarchical model of galaxy formation \citep{ledelliou05, ledelliou06}.  
As this comparison demonstrates, our luminosity function for $z=3.1$ LAEs
lies slightly below that generated by theory.  This is not surprising:
one of the key parameters of the model, the escape fraction of 
Ly$\alpha$ photons, was set using previous estimates of the density of 
$z \sim 3$~LAEs.  Unfortunately, these measurements were based on extremely 
small samples of objects, specifically, nine $z=3.1$ emitters from 
\citet{kud00} and ten $z=3.4$ LAEs from \citet{ch98}.  Since these surveys 
inferred a larger space density of Ly$\alpha$ emitters than measured in this 
paper, a mismatch between our data and the \citet{ledelliou06} models is 
neither unexpected nor significant.

\section{Star Formation Rate Density at $z \sim 3.1$}
Perhaps the most interesting result of our survey comes from a comparison
of the galaxies' Ly$\alpha$ emission with their $R$-band magnitudes.
Both quantities measure star formation rate:  Ly$\alpha$ via the combination of 
Case~B recombination theory and the H$\alpha$ vs.~star formation relation 
\begin{equation}
{\rm SFR(Ly}\alpha) = 9.1 \times 10^{-43} \, L({\rm Ly}\alpha) \ 
M_{\odot}~{\rm yr}^{-1} 
\end{equation}
\citep{kennicutt, hcm98}, and $R$, via population synthesis models of
the rest-frame UV ($\lambda 1570$)
\begin{equation}
{\rm SFR(UV)} = 1.4 \times 10^{-28} \, L_{\nu} \ M_{\odot}~{\rm yr}^{-1}
\end{equation}
\citep{kennicutt}.  If both of these calibrations hold for our sample of
Ly$\alpha$ emitters, then a plot of the two SFR indicators should scatter
about a one-to-one relation.

\begin{figure}
\figurenum{13}
\plotone{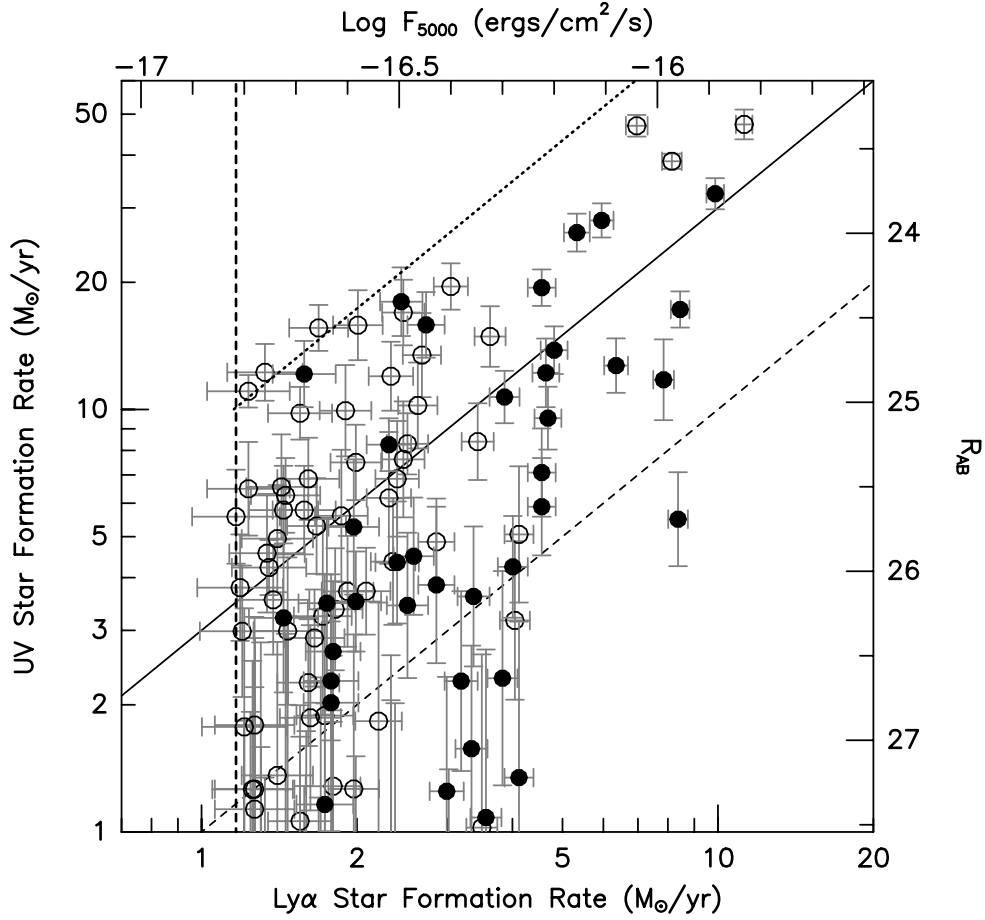}
\figcaption[sfr]
{A comparison of the star formation rates derived from Ly$\alpha$ emission
(under Case B recombination) and the UV continuum at 1570~\AA.  The solid
dots represent spectroscopically confirmed objects \citep{lira07}.  The 
diagonal dashed line shows where the two measurements are equal, while the
solid line illustrates where the UV continuum star-formation rate is three 
times the Ly$\alpha$ rate.  Our flux limit is shown via the vertical dashed
line; the approximate location of our equivalent width threshold is 
shown via the dotted line (see text).  Note that, although the two
indicators are correlated, the Ly$\alpha$-inferred rates are $\sim 3$
times less than those derived from the UV continuum.
\label{sfr}}
\end{figure}

Figure~\ref{sfr} displays this plot.  In the figure, galaxies with
Ly$\alpha$ star-formation rates less than $\sim 1.15 M_{\odot}$~yr$^{-1}$ are 
excluded by our $1.5 \times 10^{-17}$~ergs~cm$^{-2}$~s$^{-1}$ monochromatic 
flux limit, while objects with large UV star-formation rates, but weak
Ly$\alpha$ are eliminated by our equivalent width criterion.  The latter
is not a hard limit, since LAE colors range from $0 \lesssim (B+V) - R 
\lesssim 2.5$, and it is the $B+V$ continuum that is used to define 
equivalent width.  Nevertheless, if we adopt 1.4 as the upper limit on
the median color of an Ly$\alpha$ emitting galaxy (\ie\ $1 \, \sigma$ above 
the median $(B+V) - R \sim 0.65$ color of the population), we obtain the
dotted line shown in the figure.

Despite these selection effects, the Ly$\alpha$ and UV continuum star-formation
rates do seem to be correlated.  However, there is an offset:  the rates
inferred from the UV are, on average, about three times higher than
those derived from Ly$\alpha$.  While the Ly$\alpha$ SFR measurements are
generally less than $10 \, \h7^{-2} \, M_{\odot}$~yr$^{-1}$, the rest-frame
UV values extend up to $\sim 50 \, \h7^{-2} \, M_{\odot}$~yr$^{-1}$.
This discrepancy has previously been seen in a sample of 20 LAEs at 
$z = 5.7$ \citep{ajiki}, and has two possible explanations. 

\begin{figure}
\figurenum{14}
\plotone{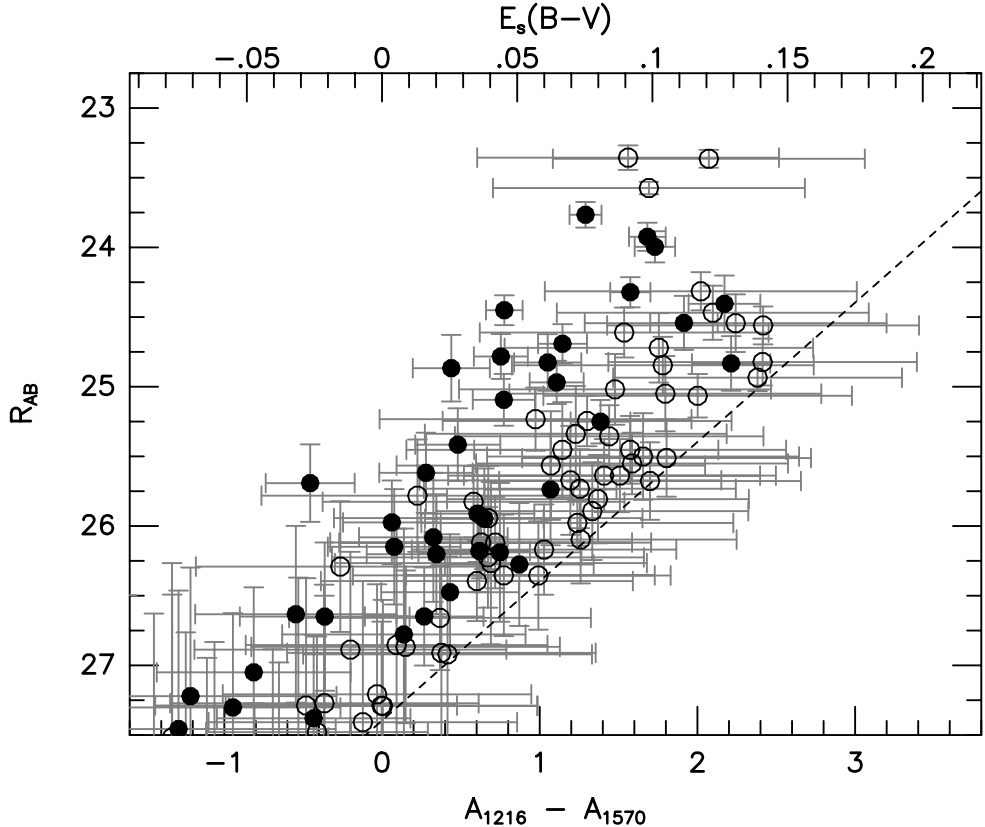}
\figcaption[extinction]
{Estimates of the internal extinction within our Ly$\alpha$ emitters,
formed using the assumption that the galaxies' ionized gas is attenuated
more than its stars \citep{calzetti}.  The solid dots represent 
spectroscopically confirmed Ly$\alpha$ emitters; the dashed line shows
the monochromatic flux limit of our survey.  Note that very little
internal extinction is needed to bring the UV and Ly$\alpha$ star-formation
rates into agreement:  even in the bright ($R > 24.5$) galaxies, the internal
extinction is never more than $E(B-V)_{\rm stars} \sim 0.1$.
\label{extinction}}
\end{figure}

The most likely cause of the offset is the galaxies' internal 
extinction.  By studying local starburst galaxies, \citet{calzetti} has shown 
that a system's ionized gas is typically attenuated more than its stars.  
In other words, while optical and IR emission-line ratios can usually be 
reproduced with a simple screen model, the shape of the UV continuum 
requires that the dust and stars be intermingled.   For a self-consistent 
solution, \citet{calzetti} suggests
\begin{equation}
E(B-V)_{\rm stars} = 0.44 E(B-V)_{\rm gas}
\end{equation}
If we apply the \citet{calzetti} law to our sample of $z = 3.1$ Ly$\alpha$
emitters, then for the UV and Ly$\alpha$ star-formation rates to be equal, 
the extinction within our LAEs must be as shown in Figure~\ref{extinction}. 
According to the figure, in most cases it only requires a small amount of dust
($E(B-V)_{\rm stars} < 0.05$) to bring the two indicators into 
agreement.   Figure~\ref{extinction} also suggests that internal
extinction becomes more important in the brighter galaxies.  This is 
consistent with observations of local starburst systems \citep[\eg][]{meurer},
and is expected if the mass-metallicity relation seen in the local universe
carries over to dust content.

Alternatively, the discrepancy between the Ly$\alpha$ and UV continuum
star-formation rates may simply be due to uncertainties in their
estimators.  Models which translate UV luminosity into star formation
rate have almost a factor of two scatter and rely on a number of
parameters, including the initial mass function and the timescale for
star formation.  The latter is particularly problematic.   Ly$\alpha$
photons are produced almost exclusively by extremely young ($< 30$~Myr), 
massive ($> 10 M_{\odot}$) stars which ionize their surroundings.  
It therefore registers the instantaneous star-formation 
occurring in the galaxy.   Conversely, continuum UV emission (at 1570~\AA) 
can be produced by populations as old as $\sim 1$~Gyr;  thus, it is a
time-averaged quantity.  If the star-formation rate in our Ly$\alpha$ emitters
has declined over time, then it is possible for UV measurements to
systematically overestimate the present day star formation
\citep{glazebrook99}.

If we assume that Ly$\alpha$ emission is an accurate measure of 
star-formation, then it is possible to integrate the Schechter function
to estimate the total contribution of LAEs to the star-formation rate
density of the $z = 3.1$ universe.  We note that this procedure does
carry some uncertainty.  If we just consider galaxies brighter than
our completeness limit ($1.5 \times 10^{-17}$~ergs~cm$^{-2}$~s$^{-1}$
or $L_{{\rm Ly}\alpha} > 1.3 \times 10^{42} \, \h7^{-2}$~ergs~s$^{-1}$)
then the star-formation rate density associated with LAEs is
$\sim 3.6 \times 10^{-3} \, \h7 \, M_{\odot}$~yr$^{-1}$~Mpc$^{-3}$, or
$1.2 \times 10^{-2} \, \h7 \, M_{\odot}$~yr$^{-1}$~Mpc$^{-3}$ if the internal
extinction in these objects is $E(B-V)_{\rm stars} \sim 0.05$.  However, 
to compute the total star-formation rate density, we need to extrapolate
the LAE luminosity function to fainter magnitudes, and even 160 objects
is not sufficient to define $\alpha$ to better than $\sim 25\%$.  
Consequently, our data admit a range of solutions.  

\begin{figure}
\figurenum{15}
\plotone{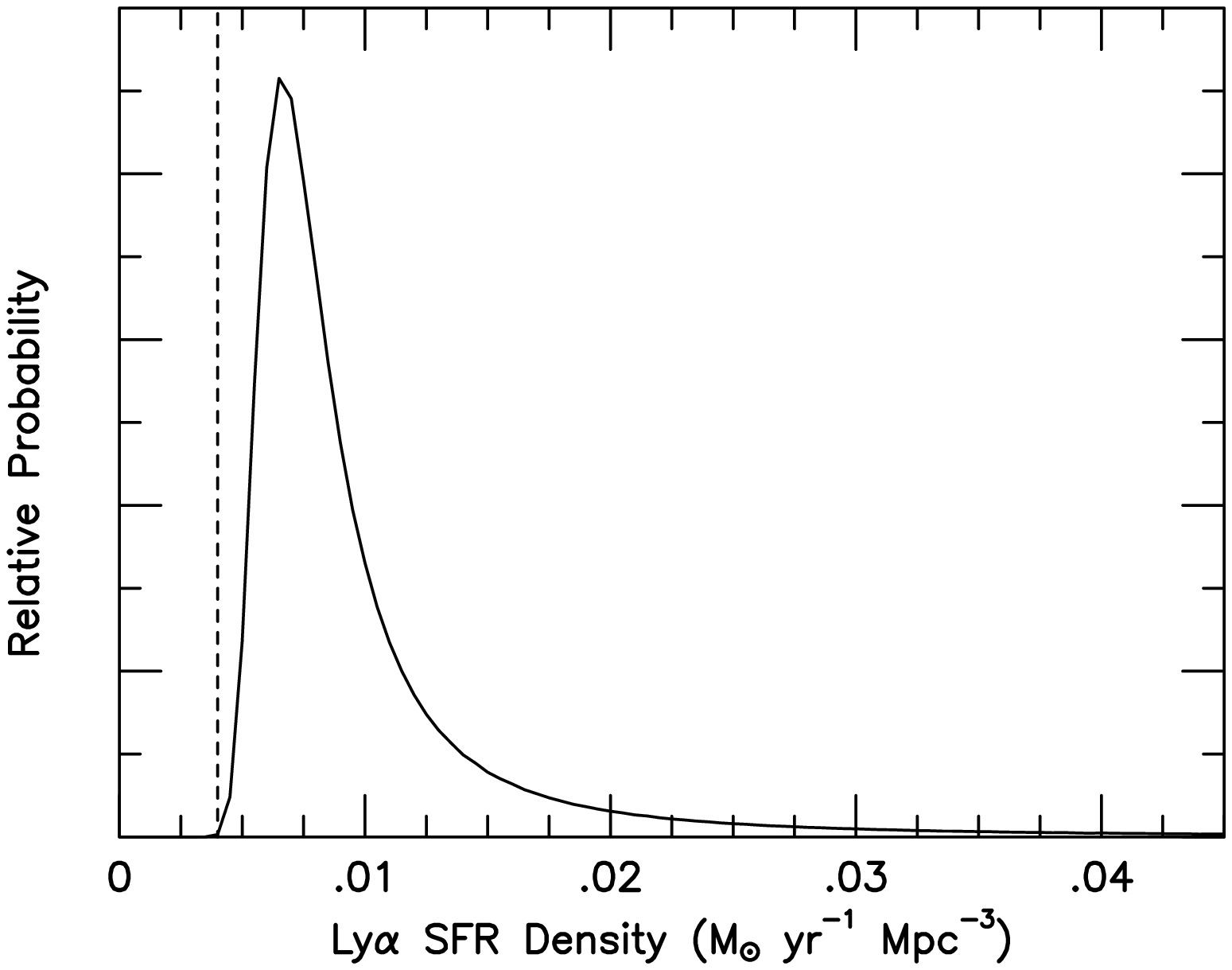}
\figcaption[sfr_prob]
{The results of our maximum likelihood analysis for the contribution 
of Ly$\alpha$ emitters to the star formation rate density of the universe.
The abscissa is the star formation rate density derived from the observed
luminosity of the Ly$\alpha$ emission line; the ordinate is the relative
probability of a solution.  The dashed line shows the {\it observed\/}
star formation rate density associated with galaxies above our completeness
limit, \ie\ without any extrapolation of the galaxy luminosity function.
No correction for internal extinction has been applied.   The figure implies
that the amount of star formation taking place in galaxies with
strong Ly$\alpha$ emission is comparable to that in Lyman-break galaxies.
\label{sfr_prob}}
\end{figure}

This is illustrated in Figure~\ref{sfr_prob}, which displays SFR 
likelihoods derived from the probabilities illustrated in 
Figure~\ref{contours}.  As the figure shows, the {\it most likely\/} value 
for the LAE star-formation rate density of the $z = 3.1$ universe 
(uncorrected for internal extinction) is 
$6.5 \times 10^{-3} \, \h7 \, M_{\odot}$~yr$^{-1}$~Mpc$^{-3}$, while
the {\it median\/} value of this quantity (defined as the point 
with equal amounts of probability above and below) is $8.6 \times 10^{-3} \,
\h7~M_{\odot}$~yr$^{-1}$~Mpc$^{-3}$.  Moreover, these numbers are
likely to be lower limits: if the discrepancy seen in Figure~\ref{sfr} is 
due to internal extinction, then the true SFR density is probably $\sim 3.5$
times higher.

The numbers above indicate that at $z = 3.1$, the star-formation rate 
density associated with Ly$\alpha$ emitters is comparable to that found
for Lyman-break galaxies.  Before correcting for extinction, our number for
the LAE star-formation rate density is $8.6 \times 10^{-3} \,
\h7~M_{\odot}$~yr$^{-1}$~Mpc$^{-3}$.  For comparison, the LBG star-formation
rate density at $z = 3.1$ (before extinction) is 
$\sim 0.01 \, \h7 \, M_{\odot}$~yr$^{-1}$~Mpc$^{-3}$ \citep{madau98, 
steidel99}.  It is true that internal extinction within Lyman-break
galaxies is typically larger than it is in our LAEs, 
$E(B-V) \sim 0.15$ \citep{steidel99}.  However, according to the 
\citet{calzetti} extinction law, the effect of dust on the emission line
flux of a galaxy is much greater than that on the stellar
continuum.  Consequently, our dust corrected SFR density for LAEs,
$\sim 0.03 \h7~M_{\odot}$~yr$^{-1}$~Mpc$^{-3}$, is $\sim 75\%$ of the
LBG value.  Of course, given the extrapolations and corrections required
to make this comparison, this number is highly uncertain.

\section{Discussion}
\begin{figure}
\figurenum{11}
\plotone{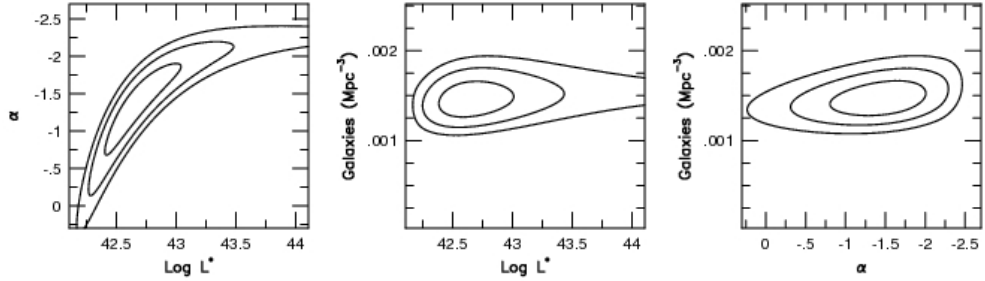}
\figcaption[contours]
{Maximum likelihood confidence contours for our \citet{schechter} function
fit to the observed distribution of Ly$\alpha$ fluxes.   The three parameters
in the analysis are the faint-end slope ($\alpha$), the bright-end cutoff
($\log L^*$) and the space density of galaxies with observed monochromatic
fluxes greater than $1.5 \times 10^{-17}$~ergs~cm$^{-2}$~s$^{-1}$,
\ie\ $L_{{\rm Ly}\alpha} > 1.3 \times 10^{42} \, \h7^{-2}$~ergs~s$^{-1}$. 
The contours of probability are drawn at $1 \, \sigma$ intervals.  
\label{contours}}
\end{figure}

The space density of $z=3.1$ Ly$\alpha$ emitters shown in Figure~\ref{contours} 
translates into a surface density of $4.6 \pm 0.4$~arcmin$^{-2}$ per unit 
redshift interval above our completeness limit.  This number is similar to 
that derived by \citet{thommes}, under the assumption that the LAE phenomenon 
is associated with the creation of elliptical galaxies and spiral bulges.
It is also consistent with the semi-analytical hierarchical structure 
calculations of \citet{ledelliou05}, though the latter predict a slightly 
larger number of $z \sim 3$ LAEs than found in this paper.  This difference is 
not significant, since the \citet{ledelliou05} models have been adjusted to 
match the previous small-volume Ly$\alpha$ surveys of \citet{kud00} and 
\citet{ch98}.  A $\sim 30\%$ re-scaling of the escape fraction of Ly$\alpha$ 
photons solves the discrepancy, and maintains the match between the 
predictions and the faint-end slope of the galaxy luminosity function.

\begin{figure}
\figurenum{9}
\epsscale{0.8}
\plotone{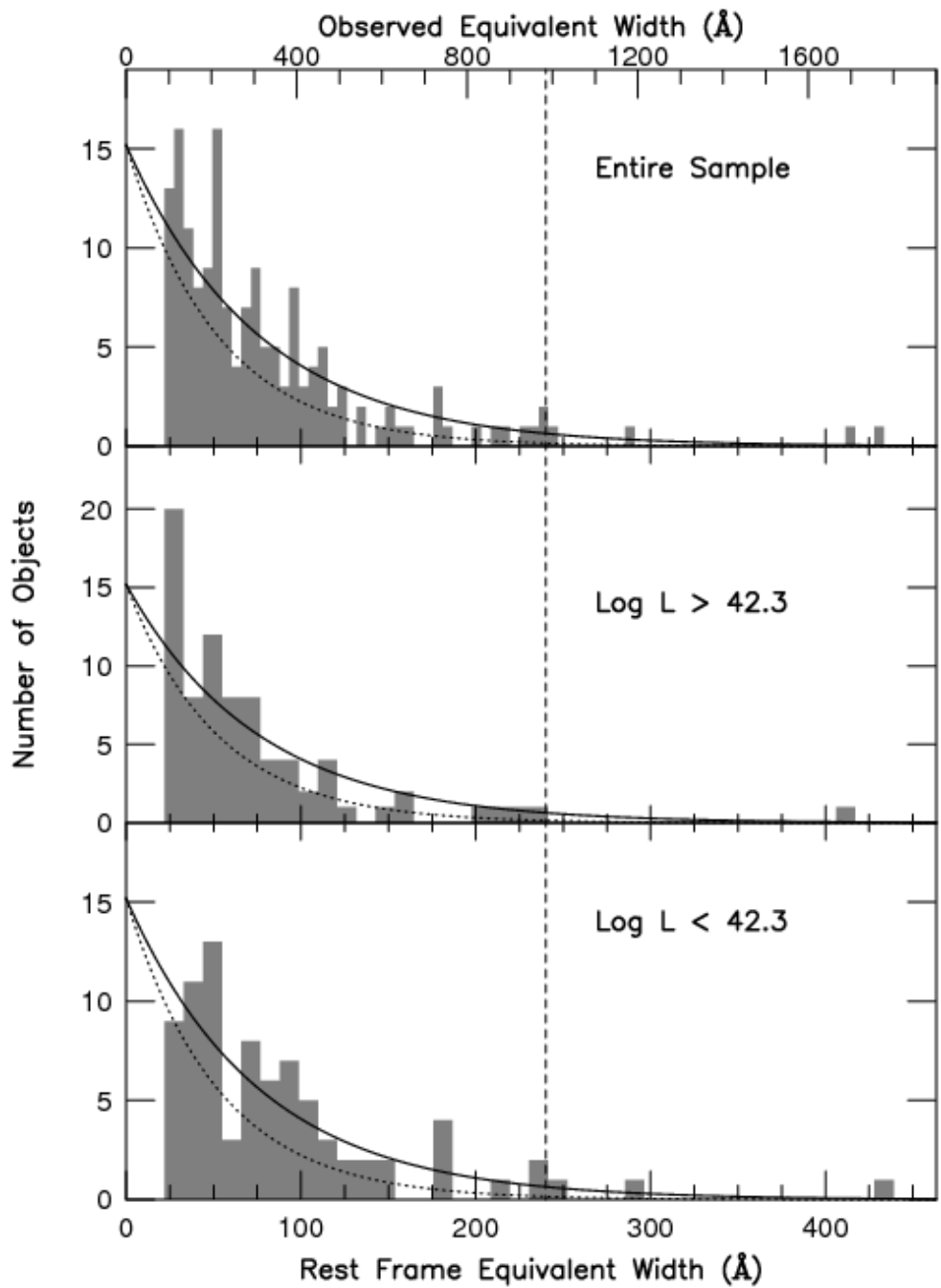}
\figcaption[ewhist]
{The top panel shows the observed distribution of equivalent widths for all the
Ly$\alpha$ emission-line galaxies in our sample.  The dotted line shows the 
apparent best-fit exponential for the distribution; the solid curve shows the 
exponential after correcting for the effects of photometric error and our
filter's non-square transmission curve.  The lower two panels divide the
sample in half, and demonstrate that the exponential law does not change 
much with galaxy luminosity.  The vertical dashed line shows the maximum 
equivalent width expected for populations with normal initial mass functions.
\label{ewhist}}
\end{figure}

More notable is the excellent agreement between the \citet{ledelliou06}
simulations and the observed distribution of Ly$\alpha$ equivalent widths 
(Figure~\ref{ewhist}).  Both are very well-fit via an exponential with a 
large ($\sim 75$~\AA) scale length.  Moreover, the models also predict that the
scale length observed for a magnitude-limited sample of galaxies (such as
that produced by the Lyman-break technique) will be much smaller than that
found via an emission-line survey.  This is consistent with the LBG results 
found by \citet{shapley}.

Nevertheless, we should emphasize that the LAEs detected in this survey are
probably not primordial galaxies in their initial stages of star-formation.
Very few of the objects have the extremely high equivalent widths calculated
for stellar populations with top-heavy initial mass functions.  More
importantly, the scatter in the galaxies' $m_{1060} - m_{1570}$ colors, 
along with the offset between the Ly$\alpha$ and UV continuum star-formation 
rates, suggests that these objects possess a non-negligible amount of dust.  
The existence of this dust argues against the Pop~III interpretation of 
$z \sim 3$ Ly$\alpha$ emitters \citep{jimenz}.

The extremely strong line emission associated with LAEs makes
these objects especially suitable for probing the evolution of galaxies and
structure in the distant universe.   The space density of $z=3.1$ emitters
shown in Figure~\ref{contours} translates into a surface density of 
$4.6 \pm 0.4$~arcmin$^{-2}$ per unit redshift interval above our 
completeness limit.  This, coupled with our measured luminosity function,
implies that in the absence of evolution, there are $\sim 12$~LAEs 
arcmin$^{-2}$ brighter than $1.5 \times 10^{-17}$~ergs~cm$^{-2}$~s$^{-1}$
in the redshift range $2 < z < 4$.  Wide field integral field units,
such as those being designed for ESO \citep{henault} and the Hobby-Eberly
Telescope \citep{virus} will therefore be able to find large numbers of 
Ly$\alpha$ emitters in a single pointing.   Moreover, because the faint-end of 
the luminosity function is steep ($\alpha \sim -1.5$), the density of LAEs 
goes linearly with survey depth.  Dropping the flux limit by a factor of 
two (to $7.5 \times 10^{-18}$~ergs~cm$^{-2}$~s$^{-1}$)
will roughly double the number of LAEs in the sample.  

With an integral-field spectrograph, it is also possible to increase the
sample of high-redshift galaxies by identifying objects with equivalent
widths lower than our detection threshold of 80~\AA\ ($\sim 20$~\AA\ in the 
LAE rest frame).   However, the gain in doing so is likely to be
small: according to Figure~\ref{ewhist}, Ly$\alpha$ rest-frame equivalent 
widths e-fold with a scale length of $\sim 75$~\AA.  If this law
extrapolates to weaker-lined systems, as suggested by the models of
\citet{ledelliou06}, then most Ly$\alpha$ emitters are already
being detected, and pushing the observations to lower equivalent widths
will only increase the number counts by $\sim 20\%$.  Furthermore, as the data
of \citet{hogg98} demonstrate, contamination by foreground [O~II] objects 
increases rapidly once the equivalent width cutoff drops below $\sim 50$~\AA\ 
in the observers frame (or $\sim 12$~\AA\ in the rest frame of Ly$\alpha$). 
Unless one can accept a large increase in the fraction of contaminants, 
surveys for high-redshift galaxies need to either stay above this
threshold, or extend to the near-IR (to detect H$\beta$ and [O~III] 
$\lambda 5007$ in the interlopers).

\acknowledgments
We would like to thank Kathy Durrell for her assistance in reducing the
data, Sean Points and Tim Abbott for their work deriving 
the transmission curve of the CTIO [O~III] interference filter, and
Cedric Lacey for providing the \citet{ledelliou06} models.
This work was supported by NSF grants 00-71238 and 01-37927 and 
HST AR10324.01A.  EG and JF acknowledge the support of NSF Astronomy \& 
Astrophysics Postdoctoral Fellowships, NSF grants 02-01667 and 03-02030.

{\it Facilities:} \facility{Blanco (Mosaic)}

\clearpage
\begin{deluxetable}{lccc}
\tablewidth{0pt}
\tablecaption{Log of Narrow-band Observations}
\tablehead{&\colhead{Exposure} &\colhead{Seeing}
&\colhead{Active} \\
\colhead{UT Date} &\colhead{(hr)} &\colhead{($\arcsec$)} &\colhead{CCDs} }
\startdata
 6 Oct 2002    &2.0    &$1.4$   &8  \\
12 Oct 2002    &1.7    &$0.9$   &8  \\
 4 Jan 2003    &2.0    &$1.0$   &8  \\
 5 Jan 2003    &3.0    &$1.0$   &8  \\
 6 Jan 2003    &3.0    &$1.1$   &8  \\
29 Nov 2003    &2.5    &$1.0$   &7  \\
 1 Dec 2003    &1.7    &$0.9$   &7  \\
23 Jan 2004    &2.3    &$1.3$   &7  \\
24 Jan 2004    &2.0    &$0.9$   &7  \\
25 Jan 2004    &2.5    &$1.1$   &7  \\
16 Feb 2004    &1.1    &$1.0$   &7  \\
17 Feb 2004    &0.8    &$1.1$   &7  \\
18 Feb 2004    &1.3    &$1.0$   &7  \\
19 Feb 2004    &1.2    &$0.9$   &7  \\
20 Feb 2004    &0.9    &$1.0$   &7  \\
\enddata
\label{log}
\end{deluxetable}
\clearpage

\begin{deluxetable}{lcccc}
\tablewidth{0pt}
\tablecaption{Candidate Ly$\alpha$ Emitters: The Statistically Complete Sample}
\tablehead{
\colhead{ID}
&\colhead{$\alpha$(2000)}
&\colhead{$\delta$(2000)}
&\colhead{Log $F_{5000}$}
&\colhead{Equivalent Width} }
\startdata
  1\tablenotemark{a}   &03:33:16.86   &$-$28:01:05.2   &$-15.596$   & 449 \\
  2   &03:33:12.72   &$-$27:42:47.1   &$-15.832$   & 392 \\
  3\tablenotemark{b}   &03:33:07.61   &$-$27:51:27.0   &$-15.860$   &  92 \\
  4   &03:32:18.79   &$-$27:42:48.3   &$-15.888$   & 251 \\
  5   &03:32:47.51   &$-$27:58:07.6   &$-15.956$   & 235 \\
  6   &03:32:52.68   &$-$27:48:09.4   &$-15.960$   & 272 \\
  7   &03:31:44.99   &$-$27:35:32.9   &$-15.972$   & 248 \\
  8   &03:31:54.89   &$-$27:51:21.0   &$-15.988$   & 310 \\
  9   &03:31:40.16   &$-$28:03:07.5   &$-16.040$   & 116 \\
 10   &03:33:22.45   &$-$27:46:36.9   &$-16.080$   & 143 \\
\enddata
\tablenotetext{a}{Candidate AGN}
\tablenotetext{b}{Foreground AGN}
\label{brightLAEs}
\end{deluxetable}
\clearpage

\begin{deluxetable}{lcccc}
\tablewidth{0pt}
\tablecaption{Candidate Ly$\alpha$ Emitters: Objects Beyond the 
Completeness Limit}
\tablehead{
\colhead{ID}
&\colhead{$\alpha$(2000)}
&\colhead{$\delta$(2000)}
&\colhead{Log $F_{5000}$}
&\colhead{Equivalent Width} }
\startdata
163   &03:33:14.82   &$-$27:44:09.1   &$-16.824$   & 380 \\
164   &03:32:08.46   &$-$27:48:43.5   &$-16.824$   & 445 \\
165   &03:33:26.22   &$-$27:46:09.0   &$-16.824$   & 468 \\
166   &03:33:11.73   &$-$27:46:51.7   &$-16.828$   & 312 \\
167   &03:31:26.49   &$-$27:50:34.3   &$-16.828$   & 209 \\
168   &03:33:05.64   &$-$27:52:47.2   &$-16.828$   & 284 \\
169   &03:31:50.46   &$-$27:41:15.2   &$-16.828$   & 364 \\
170   &03:33:17.68   &$-$27:45:44.5   &$-16.832$   & 109 \\
171   &03:31:48.98   &$-$27:53:38.7   &$-16.832$   & 322 \\
172   &03:33:10.77   &$-$27:52:41.4   &$-16.836$   & 356 \\

\enddata
\label{faintLAEs}
\end{deluxetable}
\clearpage

\clearpage
\begin{deluxetable}{cc|cc}
\tablewidth{0pt}
\tablecaption{Photometric Uncertainties}
\tablehead{
\colhead{Log $F_{5000}$} &\colhead{$\sigma$ (mag)}
&\colhead{Log $F_{5000}$} &\colhead{$\sigma$ (mag)} }
\startdata
$-15.30$    &0.022    &$-16.20$   &0.065  \\
$-15.40$    &0.024    &$-16.30$   &0.073  \\
$-15.50$    &0.026    &$-16.40$   &0.082  \\
$-15.60$    &0.031    &$-16.50$   &0.104  \\
$-15.70$    &0.033    &$-16.60$   &0.125  \\
$-15.80$    &0.038    &$-16.70$   &0.158  \\
$-15.90$    &0.042    &$-16.80$   &0.204  \\
$-16.00$    &0.050    &$-16.90$   &0.264  \\
$-16.10$    &0.058    &$-17.00$   &0.329  \\
\enddata
\label{errors}
\end{deluxetable}

\clearpage
\begin{deluxetable}{lcc}
\tablewidth{0pt}
\tablecaption{Schechter Function Parameters}
\tablehead{
\colhead{Parameter} &\colhead{Best Solution} &\colhead{Marginalized Values} }
\startdata
Log $L/L^*$ (ergs~s$^{-1}$) &42.66    &$42.64^{+0.26}_{-0.15}$ \\
$\alpha$                    &$-1.36$  &$-1.49^{+0.45}_{-0.34}$ \\
%$\phi^*$ (Mpc$^{-3}$)       &$6.02 \times 10^{-4}$ &$3.55^{+6.35}_{-1.55}
%\times 10^{-4}$ \\
$N (> 1.3 \times 10^{42} \, \h7^{-2}$~ergs~s$^{-1}$) Mpc$^{-3}$ 
&$1.46 \times 10^{-3}$ &$1.46^{+0.14}_{-0.11} \times 10^{-3}$ \\
$\phi^*$ (Mpc$^{-3}$) &$1.28 \times 10^{-3}$ &\dots \\
\enddata
\label{parameters}
\end{deluxetable}

\clearpage
\figurenum{1}
\clearpage

\pagebreak
\clearpage

\pagebreak
\clearpage

\pagebreak
\clearpage

\pagebreak
\clearpage

\pagebreak
\clearpage

\pagebreak
\clearpage

\pagebreak
\clearpage

\pagebreak
\clearpage

\pagebreak
\clearpage

\pagebreak

\clearpage
\pagebreak
\clearpage

\pagebreak
\clearpage

\pagebreak
\clearpage

\clearpage
\begin{thebibliography}{}

\bibitem[Ajiki \etal(2003)]{ajiki} Ajiki, M., Taniguchi, Y., Fujita, S.S., 
Shioya, Y., Nagao, T., Murayama, T., Yamada, S., Umeda, K., \& Komiyama, Y.
2003, \aj, 126, 2091

\bibitem[Alexander \etal(2003)]{alexander} Alexander, D.M., Bauer, F.E., 
Brandt, W.N., Schneider, D.P., Hornschemeier, A.E., Vignali, C., Barger, A.J., 
Broos, P.S., Cowie, L.L., Garmire, G.P., Townsley, L.K., Bautz, M.W., 
Chartas, G., \& Sargent, W.L.W. 2003, \aj, 126, 539

\bibitem[Blake \& Glazebrook(2003)]{blake} Blake, C., \& Glazebrook, K. 
2003, \apj, 594, 665 

%\bibitem[Bouwens \& Illingworth(2006)]{bouwens} Bouwens, R.J., \&
%Illingworth, G. 2006, \nat, in press

\bibitem[Bruzual \& Charlot(2003)]{bc03} Bruzual, G., \& 
Charlot, S. 2003, \mnras, 344, 1000 

\bibitem[Calzetti(2001)]{calzetti} Calzetti, D. 2001, \pasp, 113, 1449

\bibitem[Charlot \& Fall(1993)]{cf93} Charlot, S., \& Fall, S.M. 1993, 
\apj, 415, 580 

\bibitem[Ciardullo \etal(2002)]{c02} Ciardullo, R., Feldmeier, J.J., 
Krelove, K., Bartlett, R., Jacoby, G.H., \& Gronwall, C. 2002, \apj, 566, 784

\bibitem[Ciardullo \etal(1989)]{p2} Ciardullo, R., Jacoby, G.H., Ford, H.C., 
\& Neill, J.D. 1989, \apj, 339, 53 

\bibitem[Cowie \& Hu(1998)]{ch98} Cowie, L.L., \& Hu, E.M. 1998, \aj, 115, 1319

\bibitem[Dawson \etal(2004)]{dawson} Dawson, S., Rhoads, J.E., Malhotra, S., 
Stern, D., Dey, A., Spinrad, H., Jannuzi, B.T., Wang, J.X., \& Landes, E.
2004, \apj, 617, 707 

%\bibitem[Deharveng\etal(1994)]{deharveng} Deharveng, J.-M., Sasseen, T.P., 
%Buat, V., Bowyer, S., Lampton, M., \& Wu, X. 1994, \aap, 289, 715 

\bibitem[Jimenez \& Haiman(2006)]{jimenz} Jimenez, R., \& 
Haiman, Z. 2006, \nat, 440, 501 

\bibitem[De Propis \etal(1993)]{depropis} De Propris, R., Pritchet, C.J., 
Hartwick, F.D.A., \& Hickson, P. 1993, \aj, 105, 1243

\bibitem[Eather \& Reasoner(1969)]{eather} Eather, R.H., \& Reasoner, D.L.
1969, Appl.~Optics, 8, 227

\bibitem[Eddington(1913)]{eddington13} Eddington, A.S. 1913, \mnras, 73, 359

\bibitem[Eddington(1940)]{eddington40} Eddington, A.S. 1940, \mnras, 100, 354

\bibitem[Feldmeier \etal(2003)]{ipn2} Feldmeier, J.J., Ciardullo, R.,
Jacoby, G.H., \& Durrell, P.R. 2003, \apjs, 145, 65
 
\bibitem[Fujita \etal(2003)]{fujita} Fujita, S.S., Ajiki, M., Shioya, Y.,
Nagao, T., Murayama, T., Taniguchi, Y., Okamura, S., Ouchi, M., 
Shimasaku, K., Doi, M., Furusawa, H., Hamabe, M., Kimura, M., Komiyama, Y., 
Miyazaki, M., Miyazaki, S., Nakata, F., Sekiguchi, M., Yagi, M., Yasuda, N., 
Matsuda, Y., Tamura, H., Hayashino, T., Kodaira, K., Karoji, H., Yamada, T., 
Ohta, K., \& Umemura, M. 2003, \aj, 125, 13

\bibitem[Gallego \etal(2002)]{gallego02} Gallego, J., Garc\'ia-Dab\'o, C.E., 
Zamorano, J., Arag\'on-Salamanca, A., \& Rego, M. 2002, \apjl, 570, L1 

\bibitem[Gawiser \etal(2006a)]{gawiser} Gawiser, E., van Dokkum, P.G.,
Gronwall, C., Ciardullo, R., Blanc, G., Castander, F.J., Feldmeier, J.,
Francke, H., Franx, M., Haberzettl, L., Herrera, D., Hickey, T., 
Infante, L., Lira, P., Maza, J., Quadri, R., Richardson, A., Schawinski, K.,
Schirmer, M., Taylor, E.N., Treister, E., Urry, C.M., \& Virani, S.N. 2006a,
\apjl, 642, L13

\bibitem[Gawiser \etal(2006b)]{musyc} Gawiser, E., \etal\ 2006b, \apjs, 162, 1 

\bibitem[Gawiser \etal(2007)]{gawiser07} Gawiser, E., \etal\ 2007, in 
preparation

\bibitem[Giacconi \etal(2002)]{giacconi} Giacconi, R., Zirm, A., Wang, J.X., 
Rosati, P., Nonino, M., Tozzi, P., Gilli, R., Mainieri, V., Hasinger, G., 
Kewley, L., Bergeron, J., Borgani, S., Gilmozzi, R., Grogin, N., 
Koekemoer, A., Schreier, E., Zheng, W., \& Norman, C. 2002, \apjs, 139, 369

\bibitem[Giavalisco(2002)]{giavalisco} Giavalisco, M. 2002, \araa, 40, 579

\bibitem[Giavalisco \etal(2004)]{GOODS} Giavalisco, M., \etal\ 2004, \apjl, 
600, L93 

\bibitem[Glazebrook \etal(1999)]{glazebrook99} Glazebrook, K., 
Blake, C., Economou, F., Lilly, S., \& Colless, M. 1999, \mnras, 306, 843 

\bibitem[Hanes \& Whittaker(1987)]{hw87} Hanes, D.A., \& 
Whittaker, D.G. 1987, \aj, 94, 906 

\bibitem[Hayashino \etal(2004)]{hayashino} Hayashino, T., Matsuda, Y., Tamura, 
H., Yamauchi, R., Yamada, T., Ajiki, M., Fujita, S.S., Murayama, T., Nagao, T.,
Ohta, K., Okamura, S., Ouchi, M., Shimasaku, K., Shioya, Y., \& Taniguchi, Y.
2004, \aj, 128, 2073

\bibitem[Henault \etal(2004)]{henault} Henault, F., Bacon, R., Content, R.,
Lantz, B., Laurent, F., Lemonnier, J.-P., \& Morris, S.L. 2004, in
Proc. SPIE Vol.~5249, Optical Design and Engineering, ed.~L. Mazuray, 
P.J. Rogers, \& R. Wartmann (Bellingham: SPIE), 134

\bibitem[Hill \etal(2006)]{virus} Hill, G.J., MacQueen, P.J.,
Palunas, P., Kelz, A., Roth, M.M., Gebhardt, K., \& Grupp, F. 
2006, New Astronomy Review, 50, 378 

%\bibitem[Hill \etal(2005)]{virus} Hill, G.J., Gebhardt, K., Komatsu, E.,
%\& MacQueen, P.J. 2004, AIP Conf.~Proc.~743: The New 
%Cosmology: Conference on Strings and Cosmology, ed.~R.E. Allen,
%D.V. Nanopoulos, \& C.N. Pope (Heidelberg: Springer-Verlag), 224

\bibitem[Hogg \etal(1998)]{hogg98} Hogg, D.W., Cohen, J.G., 
Blandford, R., \& Pahre, M.A. 1998, \apj, 504, 622 

\bibitem[Hu, Cowie, \& McMahon(1998)]{hcm98} Hu, E.M., Cowie, L.L., \& 
McMahon, R.G. 1998, \apj, 502, L99

\bibitem[Jacoby \etal(1989)]{m81} Jacoby, G.H., Ciardullo, R., Booth, J., 
\& Ford, H.C. 1989, \apj, 344, 704 

\bibitem[Jacoby \etal(1987)]{jqa} Jacoby, G.H., Quigley, R.J., \& Africano,
J.L. 1987, \pasp, 99, 672 

\bibitem[Kennicutt(1998)]{kennicutt} Kennicutt, R.C. 1998, \araa, 36, 189 

\bibitem[Kodaira \etal(2003)]{kodaira} Kodaira, K., \etal\ 2003, 
\pasj, 55, L17

\bibitem[Koehler \etal(2007)]{koehler} Koehler, R.S., Schuecker, P., 
\& Gebhardt, K. 2007, \aap, 462, 7

\bibitem[Kudritzki \etal(2000)]{kud00} Kudritzki, R.-P., M\'endez, R.H.,
Feldmeier, J.J., Ciardullo, R., Jacoby, G.H., Freeman, K.C., Arnaboldi, M.,
Capaccioli, M., Gerhard, O., \& Ford, H.C. 2000, \apj, 536, 19

\bibitem[Le Delliou \etal(2005)]{ledelliou05} Le Delliou, M., 
Lacey, C., Baugh, C.M., Guiderdoni, B., Bacon, R., Courtois, H., Sousbie, 
T., \& Morris, S.L. 2005, \mnras, 357, L11 

\bibitem[Le Delliou \etal(2006)]{ledelliou06} Le Delliou, M., 
Lacey, C.G., Baugh, C.M., \& Morris, S.L. 2006, \mnras, 365, 712 

\bibitem[Lehmer \etal(2005)]{lehmer05} Lehmer, B.D., Brandt, W.N., 
Alexander, D.M., Bauer, F.E., Schneider, D.P., Tozzi, P., Bergeron, J., 
Garmire, G.P., Giacconi, R., Gilli, R., Hasinger, G., Hornschemeier, A.E., 
Koekemoer, A.M., Mainieri, V., Miyaji, T., Nonino, M., Rosati, P., 
Silverman, J.D., Szokoly, G., \& Vignali, C. 2005, \apjs, 161, 21

\bibitem[Lehmer \etal(2007)]{lehmer07} Lehmer, B.D., Brandt, W.N., 
Alexander, D.M., Bell, E.F., McIntosh, D.H., Bauer, F.E., Hasinger, G., 
Mainieri, V., Miyaji, T., Schneider, D.P., \& Steffen, A.T.
\apj, 657, 681

\bibitem[Lira \etal(2007)]{lira07} Lira, P., \etal\ 2007, in preparation

\bibitem[Madau \etal(1996)]{madau} Madau, P., Ferguson, H.C., Dickinson, M.E.,
Giavalisco, M., Steidel, C.C., \& Fruchter, A. 1996, \mnras, 283, 1388

\bibitem[Madau \etal(1998)]{madau98} Madau, P., Pozzetti, L., 
\& Dickinson, M. 1998, \apj, 498, 106 

\bibitem[Malhotra \& Rhoads(2002)]{mr02} Malhotra, S., \& 
Rhoads, J.E. 2002, \apjl, 565, L71 

\bibitem[Malhotra \& Rhoads(2004)]{mr04} Malhotra, S., \& 
Rhoads, J.E. 2004, \apjl, 617, L5 

\bibitem[Meier \& Terlevich(1981)]{mt81} Meier, D.L., \& Terlevich, R. 1981,
\apj, 246, L109

\bibitem[Meurer \etal(1995)]{meurer} Meurer, G.R., Heckman, T.M., 
Leitherer, C., Kinney, A., Robert, C., \& Garnett, D.R. 1995, \aj, 
110, 2665 

\bibitem[Monet \etal(1998)]{monet} Monet, D., Bird, A., Canzian, B.,
Dahn, C., Guetter, H., Harris, H., Henden, A., Levine, S., Luginbuhl, C.,
Monet, A.K.B., Rhodes, A., Riepe, B., Sell, S., Stone, R., Vrba, F.,
\& Walker, R. 1998, PMM USNO-A2.0: A Catalogue of Astrometric Standards
(Washington, DC: US Naval Obs.)

\bibitem[Ouchi \etal(2003)]{ouchi} Ouchi, M., Shimasaku, K., Furusawa, H., 
Miyazaki, M., Doi, M., Hamabe, M., Hayashino, T., Kimura, M., Kodaira, K., 
Komiyama, Y., Matsuda, Y., Miyazaki, S., Nakata, F., Okamura, S., 
Sekiguchi, M., Shioya, Y., Tamura, H., Taniguchi, Y., Yagi, M., \& Yasuda, N.
2003, \apj, 582, 60

\bibitem[Ranalli \etal(2003)]{ranalli} Ranalli, P., Comastri,
A., \& Setti, G. 2003, \aap, 399, 39

\bibitem[Rhoads \etal(2003)]{rhoads03} Rhoads, J.E., Dey, A., Malhotra, S., 
Stern, D., Spinrad, H., Jannuzi, B.T., Dawson, S., Brown, M., \& Landes, E. 
2003, \aj, 125, 1006

\bibitem[Rhoads \etal(2000)]{rhoads00} Rhoads, J.E., Malhotra, S., Dey, A., 
Stern, D., Spinrad, H., \& Jannuzi, B.T. 2000, \apj, 545, L85

\bibitem[Rix \etal(2004)]{GEMS} Rix, H.-W.,  Barden, M., Beckwith, S.V.W., 
Bell, E.F., Borch, A., Caldwell, J.A.R., H\"aussler, B., Jahnke, K., 
Jogee, S., McIntosh, D.H., Meisenheimer, K., Peng, C.Y., 
Sanchez, S.F., Somerville, R.S., Wisotzki, L., \& Wolf, C. 2004,
\apjs, 152, 163 

\bibitem[Schechter(1976)]{schechter} Schechter, P. 1976, \apj, 203, 297 

\bibitem[Seo \& Eisenstein(2003)]{seo} Seo, H.-J., \& Eisenstein, D.J. 
2003, \apj, 598, 720 

\bibitem[Shapley \etal(2003)]{shapley} Shapley, A.E., 
Steidel, C.C., Pettini, M., \& Adelberger, K.L. 2003, \apj, 588, 65 

\bibitem[Shimasaku \etal(2004)]{shimasaku04} Shimasaku, K., Hayashino, T., 
Matsuda, Y., Ouchi, M., Ohta, K., Okamura, S., Tamura, H., Yamada, T., \&
Yamauchi, R. 2004, \apjl, 605, L93 

\bibitem[Shimasaku \etal(2006)]{shimasaku06} Shimasaku, K., Kashikawa, N., 
Doi, M., Ly, C., Malkan, M.A., Matsuda, Y., Ouchi, M., Hayashino, T., 
Iye, M., Motohara, K., Murayama, T., Nagao, T., Ohta, K., Okamura, S., 
Sasaki, T., Shioya, Y., \& Taniguchi, Y. 2006, \pasj, 58, 313

\bibitem[Steidel \etal(1999)]{steidel99} Steidel, C.C., Adelberger, K.L., 
Giavalisco, M., Dickinson, M., \& Pettini, M. 1999, \apj, 519, 1 

\bibitem[Steidel \etal(2000)]{steidel00} Steidel, C.C., 
Adelberger, K.L., Shapley, A.E., Pettini, M., Dickinson, M., \& 
Giavalisco, M. 2000, \apj, 532, 170 

\bibitem[Steidel \etal(2003)]{steidel03} Steidel, C.C., 
Adelberger, K.L., Shapley, A.E., Pettini, M., Dickinson, M., \& 
Giavalisco, M. 2003, \apj, 592, 728 

\bibitem[Steidel \etal(1996a)]{steidel96a} Steidel C.C., Giavalisco, M.,
Dickinson, M., \& Adelberger, K.L. 1996a, \aj, 112, 352

\bibitem[Steidel \etal(1996b)]{steidel96b} Steidel, C.C., Giavalisco, M., 
Pettini, M., Dickinson, M., \& Adelberger, K.L. 1996b, \apj, 462, L17

\bibitem[Stiavelli \etal(2001)]{stiavelli} Stiavelli, M., Scarlata, C., 
Panagia, N., Treu, T., Bertin, G., \& Bertola, F. 2001, \apj, 561, L37

\bibitem[Stone(1977)]{stone} Stone, R.P.S. 1977, \apj, 218, 767

\bibitem[Taniguchi \etal(2005)]{taniguchi} Taniguchi, Y., \etal\ 2005,
\pasj, 57, 165 

\bibitem[Tapken \etal(2006)]{tapken} Tapken, C., Appenzeller, I., 
Gabasch, A., Heidt, J., Hopp, U., Bender, R., Mehlert, D., Noll, S., 
Seitz, S., \& Seifert, W. 2006, \aap, 455, 145 

\bibitem[Teplitz \etal(2003)]{teplitz03} Teplitz, H.I., Collins, N.R., 
Gardner, J.P., Hill, R.S., \& Rhodes, J. 2003, \apj, 589, 704 

\bibitem[Thommes \& Meisenheimer(2005)]{thommes} Thommes, E., 
\& Meisenheimer, K. 2005, \aap, 430, 877 

\bibitem[Thompson, Djorgovski, \& Trauger(1995)]{thompson} Thompson, D., 
Djorgovski, S., \& Trauger, J. 1995, \aj, 110, 963

%\bibitem[Tran \etal(2004)]{tran} Tran, K.-V.H., Lilly, S.J., Crampton, D., 
%\& Brodwin, M. 2004, \apjl, 612, L89 

\bibitem[Venemans \etal(2005)]{venemans} Venemans, B.P., R\"ottgering, H.J.A., 
Miley, G.K., Kurk, J.D., de Breuck, C., Overzier, R.A., van Breugel, W.J.M., 
Carilli, C.L., Ford, H., Heckman, T., Pentericci, L., \& McCarthy, P. 2005,
\aap, 431, 793 

\bibitem[Virani \etal(2006)]{virani} Virani, S.N., Treister, 
E., Urry, C.M., \& Gawiser, E. 2006, \aj, 131, 2373 

\end{thebibliography}
\end{document}